\documentclass[%
 reprint,
amsmath,amssymb,
prd,
]{revtex4-2}

\usepackage{graphicx}
\usepackage{dcolumn}
\usepackage{bm}
\usepackage{comment}
\usepackage{subfigure}
\usepackage[colorlinks=true,allcolors=blue]{hyperref}

\usepackage{url}
\usepackage{multirow}
\usepackage{amsmath}
\usepackage{amssymb}
\usepackage{amsfonts}
\usepackage{xcolor}
\usepackage{acronym}
\usepackage{ulem}
\usepackage{multirow}

\usepackage{acronym} 

\begin{document}

\newcommand{\zxt}[1]{{\textcolor{cyan}{{\sf{[ZhangXueting:#1]}}} }}
\newcommand{\nk}[1]{{\textcolor{purple}{{\sf{[Natalia:#1]}}} }}

\newcommand{\phenomd}{\texttt{IMRPhenomD}}
\newcommand{\red}{\textcolor{red}}
\newcommand{\Msun}{M_{\odot}}
\newcommand{\e}{\mathrm{e}}
\newcommand{\iu}{\mathrm{i}}
\newcommand{\tc}{t_{\mathrm{c}}}

\newcommand{\tabref}[1]{Tab. \ref{#1}}
\renewcommand{\eqref}[1]{Eq. (\ref{#1})}
\newcommand{\figref}[1]{Fig.~\ref{#1}}
\newcommand{\secref}[1]{Sec.~\ref{#1}}
\newcommand{\appref}[1]{Appendix \ref{#1}}

\newcommand{\MC}[1]{\textcolor{blue}{#1}}

\newcommand{\TRC}{TianQin Research Center for Gravitational Physics, Sun Yat-sen University, 2 Daxue Rd., Zhuhai 519082, China}

\newcommand{\TianQin}{MOE Key Laboratory of TianQin Mission, TianQin Research Center for Gravitational Physics \& School
of Physics and Astronomy, Frontiers Science Center for TianQin, CNSA Research Center for Gravitational
Waves,Sun Yat-sen University (Zhuhai Campus), Zhuhai 519082, China}

\newcommand{\SPA}{School of Physics and Astronomy, Sun Yat-sen University, 2 Daxue Rd., Zhuhai 519082, China}
\newcommand{\SUPA}{SUPA, School of Physics and Astronomy, University of Glasgow, Glasgow G12 8QQ, United Kingdom}
\newcommand{\AEI}{Max Planck Institute for Gravitational Physics (Albert Einstein Institute), D-14476 Potsdam, Germany}

\preprint{APS/123-QED}

\title{Searching for gravitational waves from stellar-mass binary black holes early inspiral}

\author{Xue-Ting Zhang}
 \altaffiliation{\TianQin}
\altaffiliation{\SUPA}
\altaffiliation{\AEI}
\email{zhangxt57@mail2.sysu.edu.cn}

 
\author{Natalia Korsakova}
 \altaffiliation{Astroparticule et Cosmologie, CNRS, Universit\'e Paris Cit\'e, F-75013 Paris, France}

 \author{Man Leong Chan}
 \altaffiliation{Department of Physics and Astronomy, University of British Columbia, Vancouver, BC V6T 1Z4, Canada}

\author{Chris Messenger}
\altaffiliation{\SUPA}

\author{Yi-Ming Hu}
\altaffiliation{\TianQin}
\email{Email:huyiming@mail.sysu.edu.cn}

\date{\today}

\begin{abstract}

The early inspiral from stellar-mass binary black holes (sBBHs) can emit milli-Hertz gravitational wave signals, making them detectable sources for space-borne gravitational wave missions like TianQin. However, the traditional matched filtering technique poses a significant challenge for analyzing this kind of signals, as it requires an impractically high number of templates ranging from $10^{31}$ to $10^{40}$.
We propose a search strategy that involves two main parts: initially, we reduce the dimensionality of the simulated signals using incremental principal component analysis (IPCA). Subsequently  we train the convolutional neural networks (CNNs) based on the compressed TianQin data obtained from IPCA, aiming to develop both a detection model and a point parameter estimation model.
The compression efficiency for the trained IPCA model achieves a cumulative variance ratio of 95.6\% when applied to $10^6$ simulated signals.
To evaluate the performance of CNN we generate the receiver operating characteristic curve for the detection model which is applied to the test data with varying  signal-to-noise ratios.  At a false alarm probability of 5\% the corresponding true alarm probability for signals with a signal-to-noise ratio of 50 is 86.5\%.
Subsequently, we introduce the point estimation model to evaluate the value of the chirp mass of corresponding sBBH signals with an error. For signals with a signal-to-noise ratio of 50, the trained point estimation CNN model can estimate the chirp mass of most test events, with a standard deviation error of  2.49 $M_{\odot}$  and a relative error precision of 0.13.


\acrodef{MBH}[MBH]{massive black hole}
\acrodef{sBH}{stellar-mass black hole}
\acrodef{MBBH}[MBBH]{massive binary black hole}
\acrodef{sBBH}{stellar-mass binary black hole}
\acrodef{CO}[CO]{compact object}
\acrodef{DWD}[DWD]{double white dwarf}
\acrodef{BBH}[BBH]{binary black hole}
\acrodef{SOBH}[]{stellar origin black hole}
\acrodef{BNS}[BNS]{binary neutron star}
\acrodef{BH}[BH]{black hole}
\acrodef{NS}[NS]{neutron star}
\acrodef{WD}[WD]{white dwarf}
\acrodef{GW}[GW]{gravitational wave}
\acrodef{EMRI}[EMRI]{extreme mass ratio inspiral}
\acrodef{CNN}{convolutional neural network}
\acrodef{SNR}[SNR]{signal-to-noise ratio}
\acrodef{AGN}{active galactic nuclei}
\acrodef{TDI}{time delay interferometry}
\acrodef{PSD}{power spectral density}

\acrodef{AK}[AK]{analytic kludge}
\acrodef{AAK}[AAK]{augmented analytic kludge}
\acrodef{NK}[NK]{numerical kludge}

\acrodef{FAP}{false alarm probability}
\acrodef{TAP}{true alarm probability}
\acrodef{ROC}{receiver operator characteristics}
\acrodef{AUC}{area under the curve}

\acrodef{ML}{machine learning}
\acrodef{ANN}{artificial neural network}
\acrodef{NN}{neural network}
\acrodef{GAN}{generative adversarial networks}
\acrodef{MMA}{Multi-Messenger Astronomy}
\acrodef{LVK}{LIGO, Virgo, and KAGRA }
\acrodef{PCA}{principal component analysis}
\acrodef{IPCA}{incremental principal component analysis}
\acrodef{PE}{point estimation}

\acrodef{LSO}{last stable orbit}

\acrodef{PN}{Post-Newtonian}
\acrodef{PSO}{particle-swarm optimization}


\end{abstract}

\maketitle


\section{Introduction}


Stellar-mass binary black hole is a typical kind of multi-band observation sources of \ac{GW}, with space-borne detectors observing the early inspiral phase of the signal, and ground-based detectors observing the late inspiral, merger, and ringdown phases. Ground-based detectors have already detected these signals, opening up the era of gravitational wave astronomy \cite{LIGOScientific:2018mvr, LIGOScientific:2020ibl,LIGOScientific:2021djp}. 
Future space-based detectors (e.g. TianQin \cite{TianQin:2015yph}, LISA \cite{LISA:2017pwj}) are likely to detect this type of source.



 
Due to the long duration ranging from months to years expected for \ac{sBBH} signals observed by space-borne detectors, they offer the potential for high-precision estimation of physical parameters, including characteristics such as the spins of the binary and its localization\cite{Marsat2020rtl,Digman:2022igm}. The information regarding the spins is particularly valuable for unraveling the evolutionary history of binary systems, given the notable variations in the effective spin predicted by various formation mechanisms\cite{2022PhRvD.106j4034T,2021PhRvD.103h3011M,Zhu:2021bpp}.Additionally, the localization of these signals can contribute significantly to constraining the Hubble constant \cite{Sesana:2016ljz,Liu:2020eko,Lyu:2023ctt,Wang:2023dgm,Huang:2023prq} 




Indeed, the detection of \acp{GW} signal from long-lived \ac{sBBH} poses a considerable challenge. 
It is important to note that the high computational cost required in practice restricts the applicable parameter area of the coherent search, taking into account both the large number of data points in each signal and a lot of templates \cite{Moore:2019pke}.
Some template-based search algorithms work well when assuming a specific range of parameters \cite{Owen:1995tm,Bandopadhyay:2023gkb}.
Owen\cite{Owen:1995tm} et al. proposed an archival search algorithm that utilizes information from ground-based detectors to narrow the parameter space for the signal being searched in space-based detector data, allowing for more efficient data analysis. However, this method relies on the gravitational wave signal being observed by both ground-based and space-borne detectors \cite{Wang:2023dgm}.
Furthermore, the semi-coherent method has been explored in zero-noise scenarios using particle-swarm optimization \cite{Bandopadhyay:2023gkb}. This was verified by a test for an sBBH injection with a chirp mass prior at a width of 2 solar masses.
If a \ac{sBBH} merger, occurring over approximately one year, can be detected by a space-borne detector, it holds the potential to offer a pre-merger warning for ground-based detectors. This necessitates the development of a swift and cost-effective pipeline for the search of \ac{sBBH} signals in the mHz frequency band. Such a pipeline would serve as a crucial link between space-borne detectors and ground-based detectors.


In recent years,  machine learning and/or deep learning algorithms have demonstrated significant potential for application in gravitational wave data analysis.\cite{George:2016hay,Gabbard:2017lja,Schafer:2022dxv,George:2017pmj,Gebhard:2019ldz,Krastev:2020skk,Huerta:2020xyq,Chua:2019wwt,Gabbard:2019rde,Dax:2021tsq,Dax:2022pxd,Bhardwaj:2023xph,Langendorff:2022fzq,2016arXiv160508803D,Ruan:2021fxq,Zhang:2022xuq,Zhao:2022qob}. 
Machine learning algorithms excel in detecting nonlinear structures within complex and long-duration signals, including events like \ac{BNS} mergers (with durations ranging from minutes to hours)\cite{Krastev:2019koe,Miller:2019jtp,Lin:2019egc,Schafer:2020kor,Yu:2021vvm,Baltus:2022pep,Aveiro:2022bkk}, continuous waves (spanning from hours to years)\cite{Bayley:2022hkz}, and extreme mass-ratio inspirals (EMRI) (spanning from a few months to years)\cite{Zhang:2022xuq}. 
As detector sensitivity improves in the future, the increasing number of events to be analyzed and challenges such as signal overlap impose distinct requirements on data analysis pipelines, such as  high computational cost. In these scenarios, machine learning algorithms demonstrate significant advantages in terms of processing efficiency and generalisation.

We propose to develop a search algorithm using \ac{CNN} for the search of \acp{GW}
from \acp{sBBH} by TianQin. Moreover, once \ac{CNN} and other neural networks complete training, their computational efficiency is very high. If we can quickly analyze the properties of a strong \ac{sBBH} from a space-borne detector, it can provide early warning to other detectors.
Of note, there are researches \cite{Lyu:2023ctt,Digman:2022igm} in which experts estimate the source parameters of an \ac{sBBH} from the actual point, presuming the data will be ready from the search phase. However, without any prior knowledge from the source, the task becomes markedly difficult. Our pipeline can operate as a first stage search at a manageable cost, helping to narrow down the parameter range and provide point parameter estimates. 



We initiated the training of \ac{CNN} using time-series data, data on the frequency domain, and data on the time-frequency spectrum.  However, the sheer volume of data points within a single signal from a \ac{sBBH}—spanning a duration of a few months—posed a challenge for \ac{CNN} training in terms of GPU memory and computational time. Consequently, we employed \ac{IPCA} to reduce the number of  data points, facilitating the construction of a search pipeline tailored for the detection of \ac{sBBH} signals.


When attempting to compress sBBH signals due to the high oscillations caused by the TianQin response, we needed to find a better signal representation to extract the dominant features. 
We have tried three different data representations for our signal: time domain, frequency domain (real and imagnary components), frquency domian(amplitude and phase). 
The amplitude is one thing we can acquire via IPCA. We undoubtedly consider the phase, but it poses a problem. The wrapped phase derived from the original responded signal also highly oscillates, and the unwrapped phase needs calibration if we consider the real data contains detector noise. In reality, noise realization will induce random artifacts for later analysis in the form of the unwrapped phase. We save the phase contribution for the future and only explore the performance with amplitude for now.


Subsequent to this, we employed the trained IPCA model
to compress the amplitude of detector data in the frequency
domain. 
Upon projecting from the IPCA model, we trained
a detection CNN model as well as a point parameter estimation CNN
model. This serves as a demonstration of the compression-plus-search pipeline methodology for the long-lasting GW search of sBBHs.



The structure of this paper is as follows: In Section \ref{sec:gw}, we introduce  \ac{sBBH} early inspiral \ac{GW}  and simulations of observations of 
\acp{GW} from \acp{sBBH} by TianQin. In section \ref{sec:method}, we explain the main techniques in data analysis that we used, including the principle component analysis and convolution neural network. In Section \ref{sec:implement}, we introduce searching strategies for \ac{GW} from \ac{sBBH}, including detection stage and point parameter estimation stage.  In Section \ref{sec:results}, we present the results. In Section \ref{sec:conclusion} 
 we summarize what we have achieved and discuss future work.

\section{GW from stellar-mass binary binary black hole inspiral}\label{sec:gw}

\subsection{Stellar-mass binary black early inspiral}



The evolution of an sBBH system comprises three distinct stages: the inspiral, merger, and ringdown phases. During each phase, the emitted gravitational waves can be characterized by various theories, including the \ac{PN} approximation, which governs the behavior of gravitational waves at different levels of accuracy.
Specifically, space borne detectors expect to capture gravitational waves radiated in the early inspiral stage of \ac{sBBH} evolution, for which the 2.5PN (post-Newtonian) approximation can be accurate enough \cite{Mangiagli:2018kpu}.

Despite that various formation channels, such as isolated binary evolution, dynamic formation, and primordial black holes, have attempted to explain the origin of \acp{sBBH}, no conclusion has been reached so far\cite{Zevin:2020gbd}. These models have made different predictions about the mass ratio, spin, and orbital parameters of the binary black holes. The validation of these predictions can be inferred via gravitational wave observations. While ground-based gravitational wave observations have completed three effective rounds of detection, more observational data are still needed to determine the model that best fits reality \cite{KAGRA:2021duu}. 
To mitigate the introduction of any preferred systematic bias in physical parameters, we incorporate a uniform distribution of \ac{sBBH} populations throughout the Universe in our simulation.

For more streamlined data analysis, we can employ \phenomd  \ \cite{Ajith:2009bn} \ to swiftly generate gravitational wave polarizations, taking into account only the dominant quadrupole moment. 
\phenomd \ includes whole evolution process of \ac{sBBH} 
(considering 3.5 \ac{PN} approximation) and can easily be used to build up our pipeline \cite{Liu:2020eko}.
In the context of the inspiral phase of the \ac{sBBH} signal, this implies that the frequency of the system satisfies $f < f_{\rm LSO}$, where $f_{\rm LSO}$ represents the frequency of the last stable orbit. Moreover, we use $h_{22}$ by \phenomd \  to characterize the original signal, expressed as:
\begin{equation} \label{h22}
    \tilde{h}_{22} (f) = A(f)\exp{(-i\Psi(f))},
\end{equation}
Here, $f$ denotes the frequency evolution range, 
The representation of the \ac{GW}  amplitude is $A(f)=\frac{\mathcal{M}^{5/6}}{\pi^{2/3}D_L}f^{-7/6}$, where $\mathcal{M} = \eta^{3/5} M$ denotes the chirp mass. The symmetric mass ratio here is $\eta=\frac{m_1m_2}{(m_1+m_2)^2}$ and the total mass is $M= m_1+m_2$ . Additionally, $\Psi(f)$ signifies the GW phase, as elaborated on in a specific set of physical parameters. These parameters include the masses $(m_1,m_2)$ of the binary system and the spins $(\chi_1,\chi_2)$ of each component within the binary, as detailed in the paper \cite{Ajith:2009bn}.

\subsection{TianQin response}

TianQin, a space-borne detector with sensitivity in the frequency range of $10^{-4}$ to 1 Hz, enabling the observation of gravitational waves originating from early inspirals of \acp{sBBH}.
Through \ac{TDI} technology\cite{Zhou:2021psj,2001CQGTinto,PhysRevD.59.102003}, we can integrate the collected data into three quasi-independent AET channels, thus effectively suppressing laser phase noise. It is common to choose these representation for TDI because the noise correlation matrix of these three combinations is diagonal.
The noise budget can be described by the following \ac{PSD}\cite{Liang:2021bde}:
\begin{align}\label{psd}
S_{A,E}(f)\ =& \frac{2}{L^2} \sin^2 f_c \left[(\cos f_c + 2) S_p(f)\right.\nonumber \\
&\left.+ 2(\cos(2f_c)+2\cos f_c+3)\frac{S_a(f)}{(2\pi f)^4}\right],\\
S_T(f)\  =& 32\sin^2 f_c\sin^2 \frac{f_c}{2}(4\sin^2 f_c S_{\rm a} +S_{\rm p}),\\
f_c =&\frac{2\pi f L}{c},\\
S_{\rm p} =&\frac{f_c^2}{L^2}\left(\frac{10^{-4}\rm Hz}{f}\right),\\
S_{\rm a}=&\left(1+\frac{10^{-4}}{f}\right)\frac{L^2}{f_c^2c^4}. 
\end{align}
Where  $ L= \sqrt{3} \times 10^8$ m is the arm length 


When the gravitational wave signal from a \ac{sBBH} system passes through TianQin the received signal will be modulated,which can be expressed  by the transfer function related to each laser links among three satellites~\cite{Marsat2018oam}:
\begin{equation}
    \tilde{s}(f) = \sum_{l} \sum_{m} \mathcal{T}^{\rm{A,E,T}}(f,t_{lm}(f)) \tilde{h}_{lm}(f).
\end{equation}


The transfer function, denoted as \((\mathcal{T}^{\rm {A,E,T}}(f, t_{lm}(f))) \), is dependent on a set of parameters, including (\(t_{\text{ref}},\phi_{\text{ref}}, \iota,\psi, \lambda, \beta\)).The reference time and phase are denoted by (\(t_{\text{ref}}, \phi_{\text{ref}} \)); the inclination and polarization angle are indicated by ($\iota$) and ($\psi$) respectively. Furthermore, the ecliptic longitude and latitude in the solar-system barycenter (SSB) are represented by ($\lambda, \beta$).

For each harmonic of the waveform, assuming it satisfies the stationary phase approximation and the shifted uniform asymptotic, its time-frequency relationship  \cite{Marsat2018oam} can be expressed as
\begin{equation}
t_{lm}(f) = t_{\rm ref } -\frac{1}{2\pi} \frac{d \Phi_{lm}(f)}{df}.
\end{equation}

The aforementioned extrinsic parameters are defined in the solar system barycentric coordinate system. By combining each harmonic \(\tilde{h}_{lm}(f)\) with its corresponding transfer function \(\mathcal{T}(f, t_{lm}(f))\), we can simulate the received signal samples of each \ac{TDI} channel. We use \phenomd  \ to describe the original waveform (only considering the 22 mode, i.e., \(l=2, m=2\) ), which is expressed as:
\begin{equation}\label{eq:signal}
\tilde{s}^{\rm{A,E,T}} =  \mathcal{T}^{\rm{A,E,T}}(f,t_{22}(f)) \tilde{h}_{22}(f).
\end{equation}

\begin{figure}
\centering
\subfigure[]{
\begin{minipage}[b]{0.42\textwidth}
\includegraphics[width=1\textwidth]
{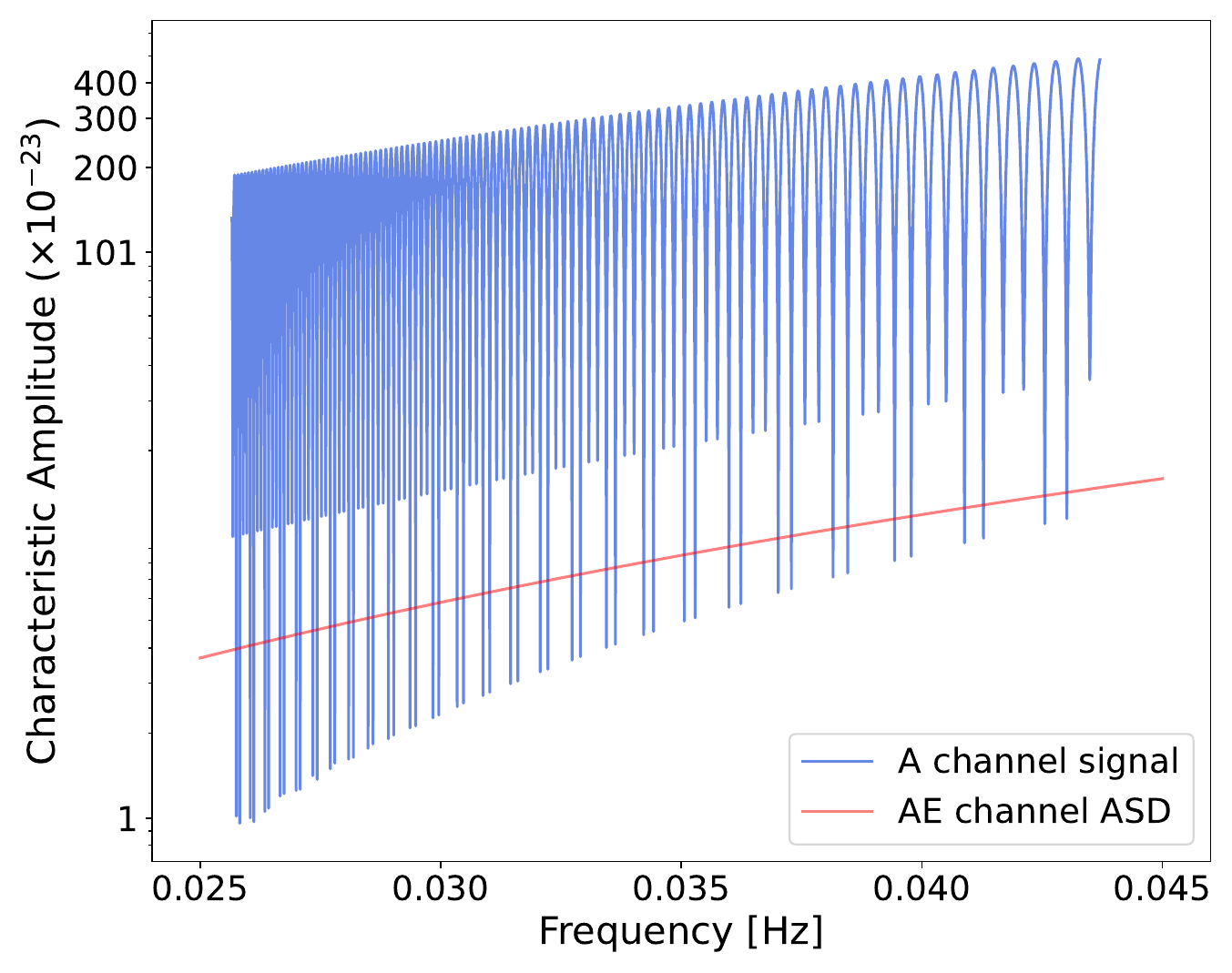} \\
\end{minipage}\label{fig:signal}
}
\subfigure[]{
\begin{minipage}[b]{0.45\textwidth}
\includegraphics[width=1\textwidth]{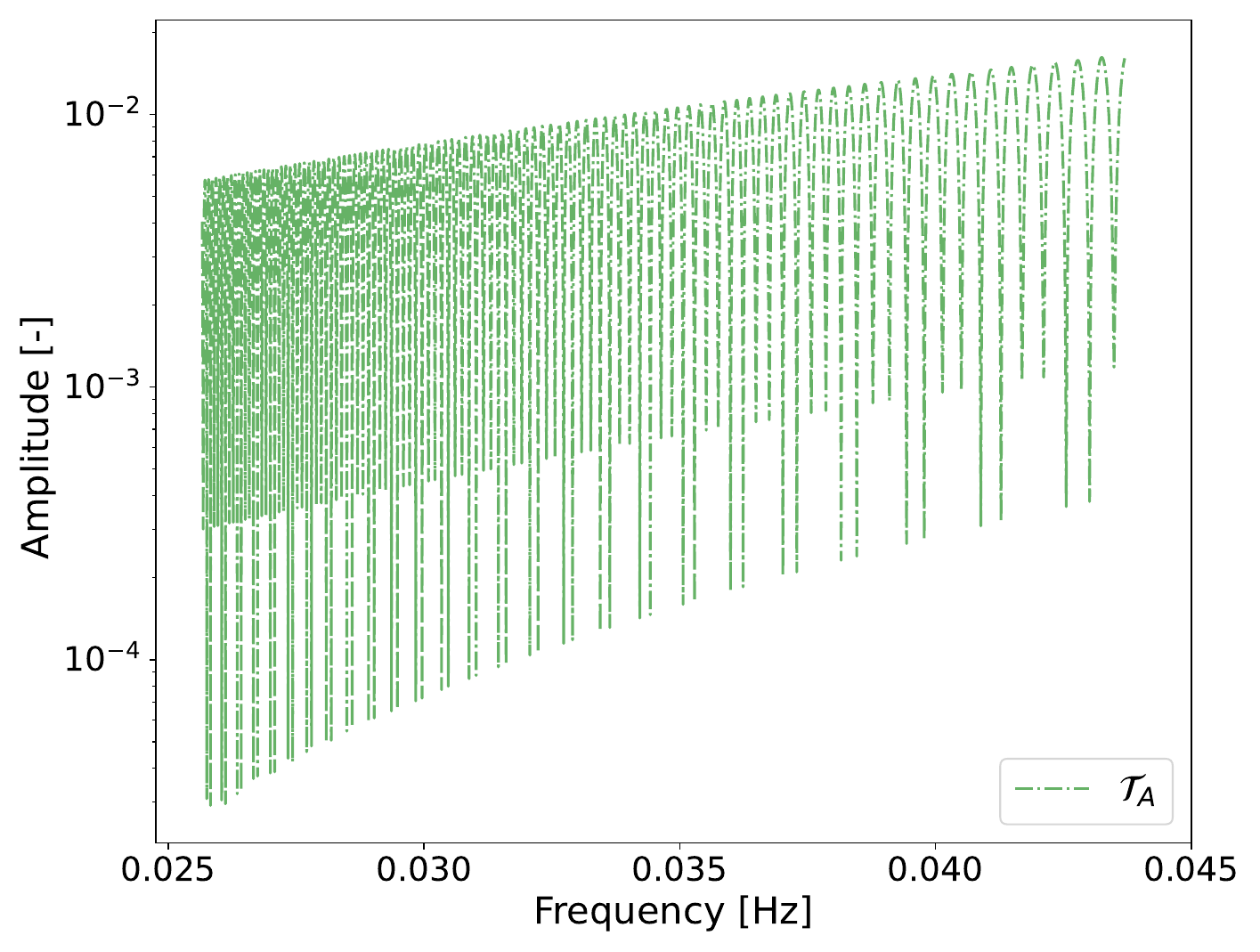} \\
\end{minipage}\label{fig:tau-A}
}
\caption{In Subfigure (a), the `A' channel signal from a GW190521-like binary black hole system is depicted as a blue curve, while the amplitude spectral density (ASD) of the `AE' channel is shown in red. The data reflects an observation period of three months at a sample rate of 0.25 Hz, with each channel containing 972001 data points. The binary system under observation comprises two black holes of approximately 85 $M_{\odot}$ and 66 $M_{\odot}$. This GW signal's \ac{SNR} is 50, and its sky location is given by ($\lambda = 5.6$ Rad, $\beta = -4.7$ Rad). Subfigure (b) presents the amplitude of the respective dimensionless transfer function in the `A' channel. The oscillation magnitude of the resultant signal is derived from this transfer function, which is governed by Equation \eqref{eq:signal}.}
\end{figure}

Using a GW190521-like \ac{sBBH} system as an example, as shown in Figure \ref{fig:signal}, this signal with the optimal \ac{SNR} of 50 exhibits multiple cycles within the sensitive band of TianQin.
The formulation for calculating the optimal \ac{SNR} is presented as follows: 
\begin{equation} 
\rho = \sqrt{\langle s|s\rangle)} 
\end{equation} 
Where $\langle \cdot | \cdot \rangle$ signifies the inner product. In the remaining context, we abbreviate optimal \ac{SNR} as SNR. 
The prominent oscillations are a result of detector modulations, primarily influenced by the transfer function.

At the same time, although we know that the detector data is composed of three channels (A, E, T), as the signal is effectively suppressed in the T channel, we only use the data from the A and E channels in subsequent analyses. 

\section{methods}\label{sec:method}


\subsection{Principle component analysis}




Taking into account the duty cycle of TianQin (operating for three months, followed by another three months off), the dimensionality of a single source in the time domain for low-frequency sBBH signals is (3, 1944000). Here, ``3" refers to the three channels of data from the \ac{TDI} AEI channel, while ``1944000" accounts for the number of data points obtained over a 3-month duration at a sampling rate of 0.25 Hz.
The use of a \ac{CNN} for processing such lengthy signals comes with its own challenges, primarily due to the large size of the signals. This poses difficulties in terms of GPU memory and training time. To tackle these issues, attempts have been made to reduce the signal size by truncating signal in each channel at the frequency domain, thus lowering the number of data points.

In this context, we employ \ac{PCA} to compress the data and extract initial features.
Using matrix decomposition,  high-dimensional data $M_{mn}$ can be represented as a set of lower-dimensional bases $U_{mk}$ 
 and the values of each basis $V_{kn}$ ~\cite{2014arXiv1404.1100S}, where $m$ represents the number of data sample, $n$ is the initial data dimension, and $k$ is the data dimension obtained through matrix transformation, with $n > k$

\begin{equation}
    M_{mn} \approx U_{mk}V_{kn} 
\end{equation}

 As $k$ approaches $n$, the product of the transformation matrix and projection obtained by \ac{PCA} is increasingly close to the matrix formed by the original data.
As the new, lower-dimensional basis ($u^i \in U_{mk}$) is chosen by selecting the eigenvalues of the covariance matrix of the data in descending order, based on their associated variance, it showcases the importance of these eigenvalues in the data representation.
In order to minimize variance loss, a larger number of components may be required. This, in turn, results in building a larger lower-dimensionality matrix $U_{mk}$. Preserving more information from the original data will reduce the variance loss, thus enabling a more accurate representation.

This method establishes the conversion matrix and realizes the dimension reduction of high-dimensional data by finding the maximum variance direction of the data to determine the first projection base, requiring the second largest variance direction to be orthogonal to the previous one, and so on, to obtain the decomposition and dimension reduction direction of the data.

Limited by the GPU memory, we cannot read a large amount of simulated signals at once. Therefore, we use \ac{IPCA} ~\cite{ipca2008} as an alternative to the actual \ac{PCA} computation process. This \ac{IPCA} method is also based on the concept of low-rank approximation, seeking a projection space similar to  \ac{PCA}, but only needing to read a batch of data at a time, instead of loading all the data into memory at once. We extracted the \ac{IPCA} model based on the third-party software \texttt{cuML} ~\cite{raschka2020machine}.

\subsection{Convolutional neural network for searching GW signals}



 \acp{CNN} are a type of deep learning algorithms that use convolutional kernels to capture and learn data features \cite{GoodBengCour16}. 
 Convolutional kernels are capable of learning to capture features at different abstraction levels. Consequently, the combination of multiple kernels allows the extraction of higher-level features, aiding the network in learning more complex patterns.
 A typical \ac{CNN} consists of convolutional layers, pooling layers and/or or fully connected layers. 
Currently, \acp{CNN} are extensively employed in the field of Gravitational Wave astronomy. 
They are utilized for various purposes,
such as detecting \ac{GW} signals
\cite{George:2016hay,George:2017pmj,Gabbard:2017lja,Verma:2022zsj} 
and classifying glitches
\cite{George:2017qtr,Alvarez2023dmv}.
The convolution kernels of these networks play a crucial role in characterizing the distinctive features of the signals and glitches.

\subsubsection{Signal detection}
In the task of signal detection, detector data can be modeled as the data that either contains a signal or not.
Using a CNN for signal detection means classifying the detector dataset, which is essentially calculating the probability that the data contains a signal. The specific mathematical model can be represented as:
\begin{align}
    y_{\rm pred} &=  {\mathcal{P}^{\rm Detection}_{\rm CNN} (x)}, \\ \notag
    x&=g(d_f). \\ \notag
    d_f  &=\left\{
\begin{aligned}
& s_{f}(\theta) + n_f\\
& n_f.\\
\end{aligned}
\right.
\end{align}
where $x$ is the input to a \ac{CNN}, which also can be a suitable representation  $g$ of the detector output data $d$. In our case, we take the compressed amplitude of the simulated data $d_f$ in  the frequency domain as the input to \acp{CNN}. $y_{\rm pred}$ is the probability  whether this input contains a \ac{GW} signal or not.

This detection task is equivalent to using a \ac{CNN} for  classification. 
In this scenario, the data is categorized into two  classes with distinct labels. In our case the detector data containing signals is assigned the label 1, while the detector data with pure noise is labeled as 0.
We can train the CNN based on cross-entropy. The formula for binary cross-entropy is as follows:
\begin{align}\label{eq:bce_loss}
\rm{Loss_{bce}}  =  -\frac{1}{N} \sum^{i=N}_{i=1}& \left[ y_i \cdot \log(y^i_{\rm pred}) + \right.\nonumber\\
&\left. (1 - y_i) \cdot \log^i (1-y^i_{\rm pred})\right]  
\end{align}
where $y_i$ represents the input label and $y_{\rm pred}$ is the corresponding output probability of the CNN.

\subsubsection{Point estimation}

Following our successful predictions regarding the presence of a signal in the data, we can proceed to evaluate the parameters of the \ac{GW} signal which is present in the data. 
Consequently, we employ the similar \ac{CNN} architecture that has proven effective in the detection task, repurposing its structure to estimate 
the physical parameters of the signal — a task 
that is commonly referred to as `regression' in machine learning applications. This approach will yield point estimates of the physical parameters
The specific mathematical model can be represented as:
\begin{align}
   \mathbf{\theta_{\rm pred}}  &= \mathcal{P}^{\rm Estimation}_{\rm CNN}  (x), \\ \notag
    x&=g(d_f(\theta_{\rm actual})). \\ \notag
    d_f(\theta) &= s_f(\theta) +n_f.
\end{align}
In this formula, $d_f(\theta)$ is a sample of signal plus noise in  the frequency domain, and $x_i$ is the representation of the data sample. There is a certain function mapping relationship between the input and physical parameters. A well-trained CNN model can learn this mapping $f_{\rm CNN}$ .

We use the Mean Squared Error (MSE) to train the difference between the input chirp mass value ${\theta_{\rm actual}}$ and the \ac{CNN} output estimate $\theta_{\rm pred}$, thereby constructing a one-to-one mapping relationship between the predicted physical parameters and the corresponding input data. The formula for the mean squared error is as follows:
\begin{equation}\label{eq:mse_loss}
    \rm{Loss_{mse} =  \frac{1}{N}\sum_{i=1}^{N} (\theta_{\rm {actual}}^i - \theta_{\rm pred}^i)}
\end{equation}
where $N$ is the number of the used data, $\theta_{\rm actual}$ is the actual physical parameter set of the corresponding data $d(\theta)$, and $\theta_{\rm pred}$ is the point estimation of the physical parameter set.


\subsubsection{CNN Architecture}


The final \ac{CNN} structure used in signal detection is shown in Table \ref{tab:CNN_architecture}. The final output layer of the \ac{CNN} is augmented with the softmax function to normalize the output, enabling its interpretation as the probability of whether the data contains a \ac{GW} signal. The other layers utilize the `ReLU' activation function to achieve non-linear mapping from the input to the output. The `ReLU' function yields zero for any negative input while returning positive values without any changes, thereby introducing sparsity into the model and potentially improving computational efficiency in the signal detection task.



\begin{table}[ht]\footnotesize
\caption{\label{tab:CNN_architecture}The architecture of the \ac{CNN}. The number of training parameters is 497466. The stride for each layer is 1.
}
\centering
\begin{ruledtabular}
\begin{tabular}{lcp{0.1\textwidth}<{\centering}lcp{0.1\textwidth}}
& \textbf{ Layers} 
& \textbf{ Neutrons number}
& \textbf{Kernel size}  
& \textbf{Output size}\\
\hline
        1 & Input       & $(2 \times 480)$  
        & $\ldots$
        & $\ldots$
        \\   
        2 & Convolution &64 & (  $1 \times 5$ ) & (  $64 \times 96$ ) \\ 
        3 & BatchNorm    &64  
        & $\ldots$
        & (  $64 \times 96$ )\\ 
        4 & Convolution &128 
        &( $1 \times 5$ ) &  (  $128  \times 92$ )  \\ 
        5 & Convolution &256 &
        ( $1 \times 5$ )&  (  $256 \times 88$ )  \\ 
        6 & Convolution &64
        &( $1 \times 3$ ) 
        &(  $64 \times 86$ )  
        \\
        7 & Convolution &16
        &( $1 \times 3$ )
        &  (  $16 \times 84$ ) 
        \\
        8 & Convolution &8
        &( $1 \times 3$ ) &  (  $8 \times 82$ )  \\
        9 & Flatten & $\ldots$ 
        &
        &   656 \\ 
        10 & Dense &$\ldots$ 
        &
        &   64 \\ 
        11 & Dropout & p = 0.5 
        & $\ldots$
        & \\
        12  & Dense &$\ldots$
        &
        &  32 \\
        13 & Dropout & p = 0.5 
        & $\ldots$
        & \\ 
        14 &  Output &$\ldots$ 
        &
        &  2  \\
\end{tabular}
\end{ruledtabular}
\end{table}

In the task of point estimation, we employ a CNN architecture similar to that used for signal detection. To elaborate, firstly, all `ReLU' activation functions are replaced with the `Tanh' function, and we remove any activation function from the output layer. 
This is done because for point estimation we noticed that `ReLU' function did not yield sufficiently good results therefore we replaced with a smooth function.
The `Tanh' function maps real-valued numbers to a range between -1 and 1, essentially compressing the inputs into a narrower range, which can facilitate convergence during training in our case.
Secondly, we did not use two dropout layers in the point estimation task, as this would discard too many features. This strategy allows a more sophisticated CNN model to handle a more challenging point parameter estimation task than the signal detection task.




\section{Implementation}\label{sec:implement}

To obtain a well-trained machine learning model, the representation of data and labels are crucial.  Initially, the data from each output channel is subjected to a whitening operation, where it is divided by its respective noise's \ac{PSD}, as demonstrated in Equation \eqref{psd} . Following this process, the data is then fed into the \ac{CNN} model.
However, facing the situation where sBBH has many individual data points, this directly challenges the size of the GPU memory and the training duration. Therefore, we have employed the \ac{IPCA} method to compress the data, obtain a new representation of observation data, and then use the newly compressed features to train a \ac{CNN} to differentiate whether the original observation data contains a \ac{sBBH} signals. In cases where signals are included, we further train the similar CNN structure to estimate the chirp mass of the candidate signal. 

\subsection{Data simulation}

\begin{table*}[!htp]
\caption{Physical parameters and their meanings: When considering \ac{sBBH} as observed by space-borne detectors, it's worth noting that during the early inspiral stage, which spans several months, the time to coalescence is approximately set at one year, ranging from 9 months to 15 months.}
\centering
\begin{tabular}{|l|l|l|l|}
\hline        \textbf{Notation} & \textbf{Meaning} & \textbf{Distribution }  & \textbf{Units}\\
\hline
	%
       $m_1$  &  primary mass &  $\log$ uniform [5,100] & $M_{\odot}$ \\ \hline
        $ m_2$  & secondary mass & $\log$  uniform [5,$m_1$]  & $M_{\odot}$\\ \hline
        $\chi_{1z}$ &  dimensionless aligned-spin magnitude of component 1 & uniform [-1,1) & -\\ \hline
       $\chi_{2z}$ & dimensionless aligned-spin magnitude of component 2 & uniform  [-1,1) & - \\ \hline
        $\lambda$  & the ecliptic longitude of the source & uniform  [$0,2\pi$] & Radian\\ \hline
        $ \beta_S$  & the ecliptic latitude of the source & $\sin \beta = ( \rm{uniform [-1,1]}) $ & Radian\\ \hline
$\iota$ & inclination & $ \cos \iota = ( \rm{uniform  [-1,1]})$ & Radian \\ \hline
$\psi$  & polarization &  uniform [$0,\pi$] & Radian\\ \hline
        $\Phi_c$  & merger phase & uniform [$0,2\pi$] & Radian \\ \hline
        $t_c $ & merger time at the reference time $t=0$ s   &  uniform [23760000,39312000]& Seconds \\ \hline
         \multirow{2}{*}{$D_L$}   &  \multirow{2}{*}{luminosity distance to source} &  re-scaled inversely by the SNR 50,   & \multirow{2}{*}{Mpc}\\
 & &the reference distance is 50 Mpc. & \\ \hline
        $f_0$ & initial frequency at the reference time $t=0$ s & calculated by Equation \eqref{eq:frequecy-of-time} & Hz\\
        \hline
\end{tabular}
\label{tab:IMRPheD-parameters-eng}
\centering
\end{table*}

We strive to explore technical possibilities within a baseline distribution of \ac{sBBH} populations. Specifically, we consider simulating signals uniformly distributed in the co-moving volume, shown in Table \ref{tab:IMRPheD-parameters-eng}.
We have utilized \phenomd ~ waveform  to generate  signals in the frequency domain \cite{Ajith:2009bn}, with parameters range set as shown in the Table  \ref{tab:IMRPheD-parameters-eng}, and analyze them buried in Gaussian noise, assuming the TianQin \ac{TDI} AET PSD \eqref{psd} \cite{Liang:2021bde} .

Subsequently, we conduct pre-processing on the simulated detector data, including whitening operations. 


\begin{figure}[!htp]
\centering
\includegraphics[width=0.45\textwidth]
{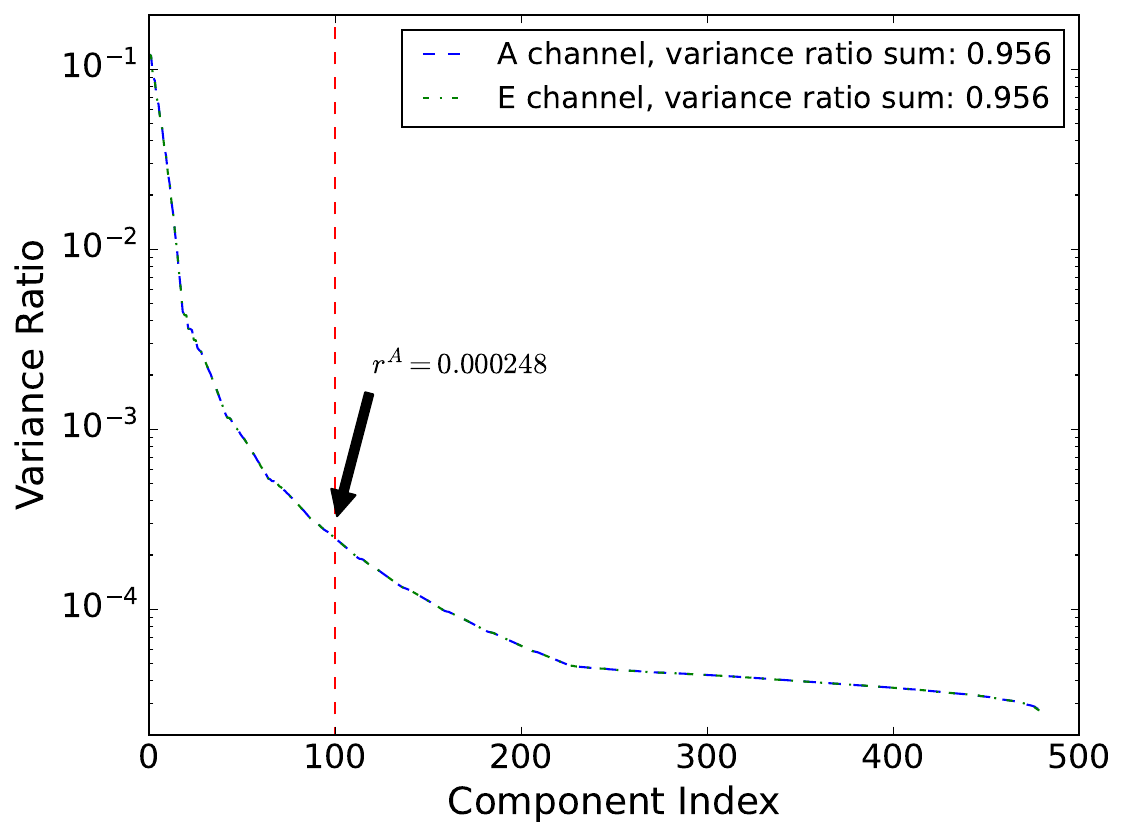}
\caption{The variance ratio of each component is arranged in descending order.
The cumulative variance of $k$ components across the entire dataset amounts to 95.6\% in each IPCA model. Because A and E channel signals are orthogonal, it is unsurprising that both of the extracted IPCA models have the same variance ratio for each component. 
The 100th component has a variance ratio of 0.000248  in $\rm{IPCA}^{A}$ model, which is less than $10^{-3}$.The cumulative variance ratio of the preceding 100 components is 93.2\%, indicating that the  subsequent components contribute less to the feature representation.}
\label{fig:variance}
\end{figure}

\begin{figure*}[!htbp]
\centering
\includegraphics[width=0.9\textwidth]{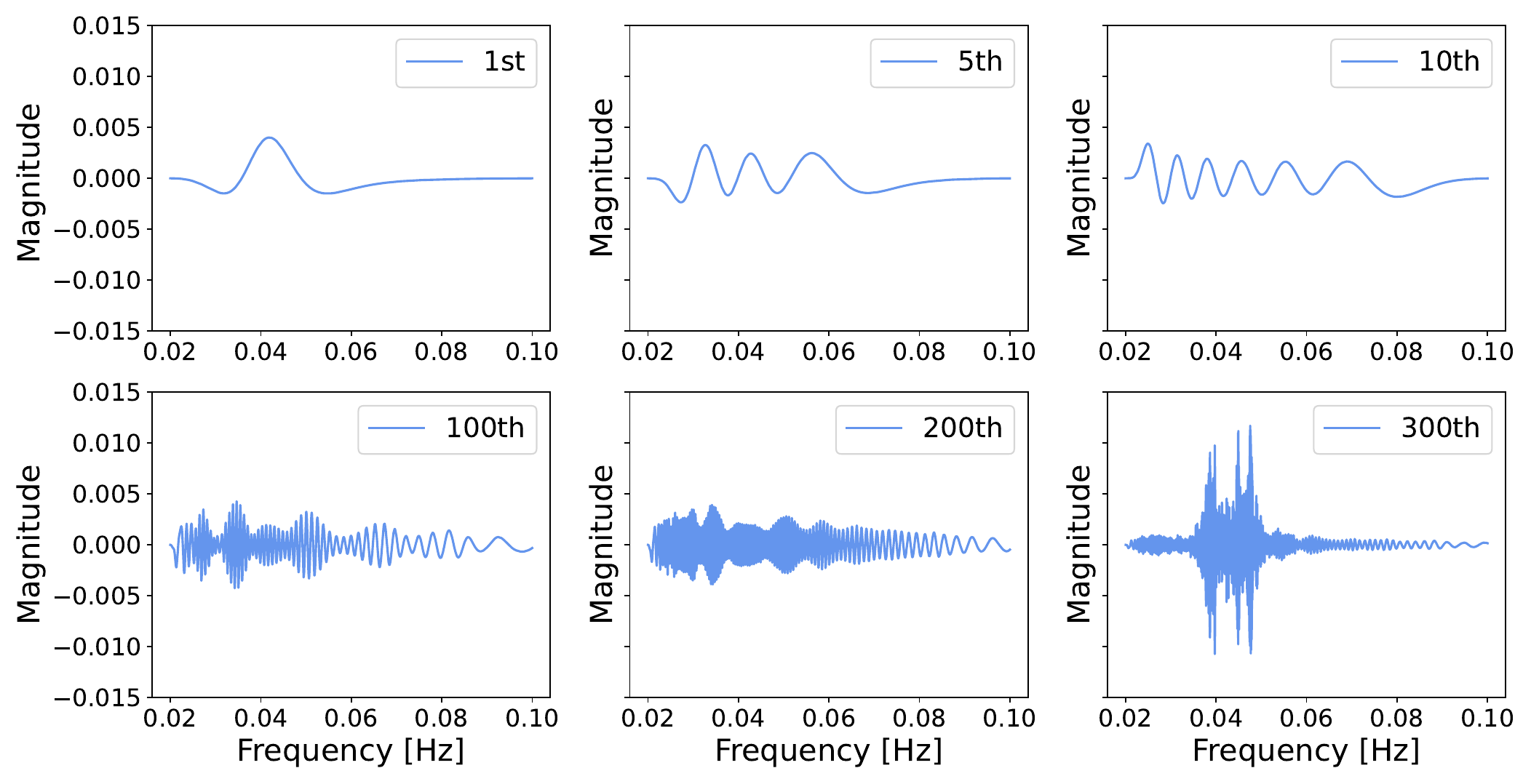}
\caption{
Each eigenvector in the IPCA serves as a basis.
The preceding few bases can effectively capture the primary signal evolution, which can be described by the Post-Newtonian approximation \ref{h22}, depending on the chirp mass. 
Higher-order bases can capture multiple oscillation features, primarily originating from the response of the TianQin detector.
} 
\label{fig:basis}
\end{figure*}

\begin{figure}[!htp]
\centering
\includegraphics[width=0.45\textwidth]
{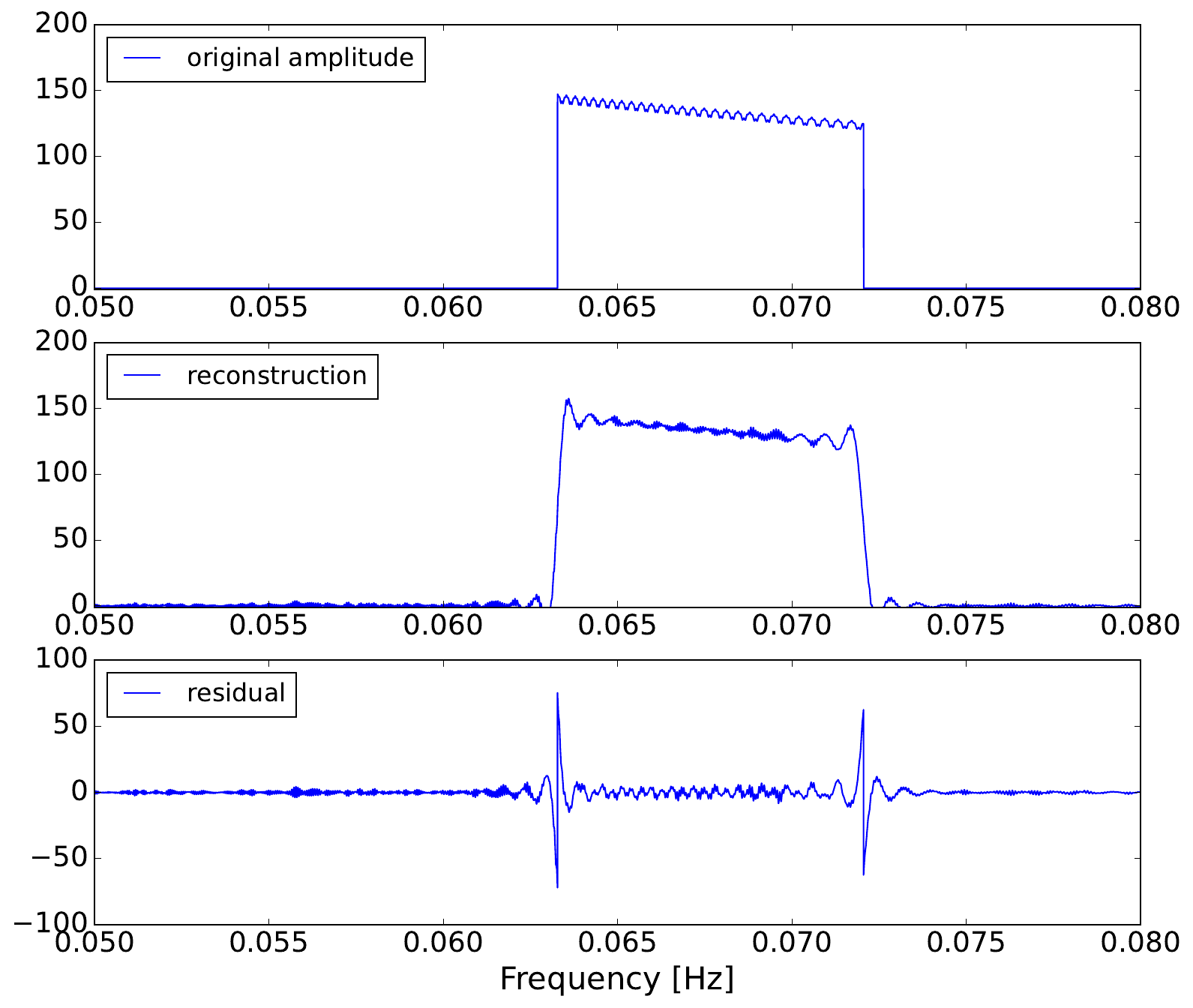}
\caption{The original amplitude from a random signal in `A' channel with \ac{SNR} 50 and the reconstruction from ${\rm IPCA}^{\rm A}$.The system in question has a chirp mass and symmetric mass ratio of approximately 8.3 $M_{\odot}$ and 0.18 respectively, and coalesces in a time-span of 28,923,738 seconds. Additional source parameters are referenced in Table \ref{tab:IMRPheD-parameters-eng}, including ( $m_1=18.05 \ M_{\odot}$, $m_2=5.4\ M_{\odot}$, $\cos (\iota)= 0.1$, $\psi=2.74 $ Rad, $\phi_c=2.64 $ Rad, $\chi_{1z}=-0.36$, $\chi_{2z}=0.59$, and sky position at $\lambda=2.07$ Rad,$\beta=-0.63$ Rad.)} 
\label{fig:rec}
\end{figure}

\begin{figure*}[!htp]
\centering
\includegraphics[width=0.7\textwidth]
{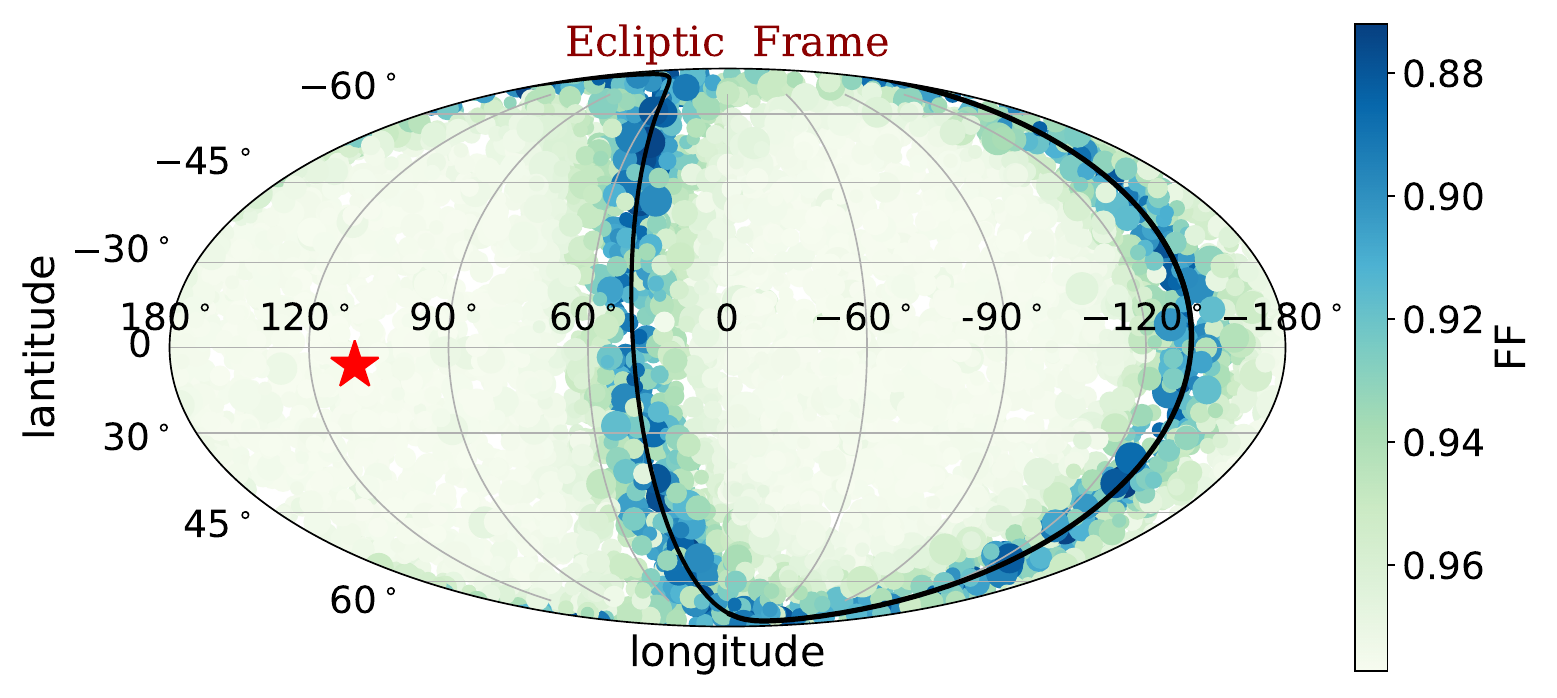}
\caption{A sky map showing the amplitude fitting factors for gravitational wave signals from sources distributed across the sky . In the map, the red star symbolizes the position of the \ac{DWD} system J0806, which aligns with the direction of the TianQin constellation. The black solid curve delineates the orbit of TianQin. Darker blue points indicating smaller amplitude fitting factors, suggesting a lower similarity between the reconstruction and the original signal. Additionally, the size of each point corresponds to the chirp mass of the source.  It's worth noting that sources located on TianQin's orbit tend to demonstrate lower amplitude fitting factors.\label{fig:ff-skylocation-map}}
\end{figure*}

\subsection{Compression}\label{subsec:pre-process}




We adopt \ac{IPCA} to extract the principal components from the signals, thereby reducing the dimensionality. We tested on different data representations, including time-domain signals, real and imaginary parts of frequency-domain signals, amplitude and phase of frequency-domain signals, etc.
In our evaluations of various representations of  responded  signals, it has been observed that \ac{IPCA} encounters challenges in extracting features from signals exhibiting substantial fluctuations. Consequently, we have chosen to exclusively derive the \ac{IPCA} model from the amplitude of frequency-domain signals. This yields the corresponding $\rm IPCA^i$, where `i' notes the detector output channel, designated as either A or E.
We used the IPCA method from the \texttt{cuML} ~\cite{raschka2020machine} third-party package to load signals into GPU memory in batches, thereby extracting the principal components of signals from a single channel after applying the detector response. These principal components represent the main characteristics of the signal amplitude.



Taking the extraction process of the amplitude ${\rm IPCA}^{\rm A}$ model from the \ac{TDI} A channel signal as an example, the size of the amplitude of a single signal in the A channel is (1,972001) obtained over a 3-month duration at a sampling rate of 0.25 Hz. By sampling within the astrophysical parameter range in Table \ref{tab:IMRPheD-parameters-eng}, we  generate 4800000 simulated signals and a set of amplitudes, with the dimensionality is   $(4800000 \times 1 \times 972001)$. 
The E channel signals will undergo a similar process for the extraction of the ${\rm IPCA}^{\rm E}$ model, maintaining an equivalent training set size as the A channel signal's compression, represented by $(4800000 \times 1 \times 972001)$.

The variance ratio, denoted as $r_j$ in each IPCA model, signifies the variance $\sigma_j$ of the $j$-th basis in relation to the total variance  $\Sigma$ of all signals, given by 
\begin{equation}
r_j = \frac{\sigma_j}{\sum}.
\end{equation}
These ratios are arranged in descending order, reflecting the efficacy of the extraction process. A lower index indicates a greater variance direction for the basis. 
As more basis components are considered, the cumulative variance ratio increases, thereby minimally affecting the compression loss.


In our case, amplitude fitting factor is a better measure to the performance of the component extraction.
We obtain the projection value of a \ac{GW} signal amplitude in a channel, based on the corresponding \ac{IPCA} principle components or eigenvectors.
By making use of the amplitude projection values (equivalent to the coefficients of the principal components),  
we can reconstruct its amplitude of the original signal and then measure the similarity between the reconstruction and the original signal amplitude by calculating the amplitude fitting factor. The formula for calculating the amplitude fitting factor is shown below.
\begin{equation}
    \rm{FF} =  \frac{\langle \left\vert {s}({\theta})\right\vert |  \left\vert s'({\theta}) \right\vert \rangle}{\sqrt{\langle \left\vert s({\theta}) \right\vert |  
\left\vert s({\theta}) \right\vert \rangle} \sqrt{\langle 
\left\vert s'({\theta}) \right\vert |  
\left\vert s'({\theta}) \right\vert \rangle}}. 
\end{equation}
Here, $s({\theta})$ is the original responded signal, $s'({\theta})$ is the reconstruction of the respective IPCA model, while $\langle \cdot | \cdot \rangle$ denotes the inner product between two signals, but calculated solely by amplitude.

\begin{align}
    &\langle 
   \left\vert s({\theta}) \right\vert |  
   \left\vert s'({\theta})
   \right\vert \rangle   \nonumber \\  
&=  4 {\int_{f_{\rm min}}^{f_{\rm max} } \frac{\left\vert s({\theta}) \right\vert \cdot \left\vert s'({\theta})
   \right\vert}{S_{n}(f)}   {\mathrm d}f}    \nonumber 
 \\ 
   & \geq 4 {\int_{f_{\rm min}}^{f_{\rm max} } \frac{\left\vert s({\theta}) \right\vert \cdot \left\vert s'({\theta})
   \right\vert \cdot \cos \Phi}{S_{n}(f)}   {\rm d} f} \nonumber \\ 
    &= \langle 
    s({\theta})  |  
   s'({\theta})
   \rangle ,   
\end{align}
where $f_{\min} = 0.02 $ Hz and $f_{\max} = 0.1 $ Hz. $\Phi$ is the difference between two signals, and $-1 \leq \cos \Phi \leq 1$.

As we analyze a multi-band \ac{GW} event with a binary coalescence time of approximately 1 year and a duration of 3 months, the frequency range of most signals should be within the band from 0.02 Hz to 0.1 Hz, as calculated by the instantaneous frequency formula at the time \(t\) before the merger time \(t_c\) :
\begin{equation}\label{eq:frequecy-of-time}
f=\left( \frac{5}{256} \right)^{3/8} \mathcal{M}^{-5/8}\left( t_c-t \right)^{-3/ 8}.
\end{equation} 
Here, \(\mathcal{M}\) represents the chirp mass of the binary system.
Therefore, we truncated the amplitude of the frequency-domain signal, reducing the size of each responded channel's signal from 972001 to 622080.



Finally, we trained the each \ac{IPCA} model using 4800000 simulated signals, yielding a transformation matrix of (622080, 480).  The training took about 57 hours using a GPU with A100 32GB memory. 
Employing around
1,000,000 training samples presents a reasonable compro-
mise as indicated in Appendix. 
 The cumulative variance ratio of these 480 components in ${\rm IPCA}^{\rm A}$ amounts to 0.958, indicating that data compressed through IPCA retains most of the variances from the amplitude of GW signals, as shown in Figure \ref{fig:variance}.
Each component is a basis vector visually represented  in Figure \ref{fig:basis}.
From these 480 bases, it's apparent that all are focused on the high fluctuations in  relatively lower frequency band, rather than on the high frequency. 
Using these \ac{IPCA} model, the simulated data from a single detector (2 x 9720001) can be compressed into a matrix of size (2, 480), achieving a compression factor of the order of 1000. 
This process can also be referred to as the data pre-processing module.
In the Figure \ref{fig:rec}, an example is depicted within the considered frequency band. This illustration provides a comparison between the original signal amplitude and the reconstructed signal amplitude, based on estimated ${\rm IPCA}^{\rm A}$ components, together with their residual. We can see that boundaries at the signal's cutoff frequencies have higher residuals which means greater loss resulting from the compression process. This probably happens due to the sharp edges of the original signal.

To demonstrate the compression loss, we examine the amplitude fitting factor across various astrophysical parameters. 
Lower amplitude fitting factors predominantly correlate with high oscillations in the signal amplitude, derived from TianQin's response. This response is represented by  transfer functions that relies on 6-dimensional parameters, including the sky location. In Figure \ref{fig:ff-skylocation-map}, the sources in specific locations (like the source localization in Figure \ref{fig:tau-A} ), related to TianQin's orbit, have a tendency to display high oscillation. This behavior subsequently leads to a lower amplitude fitting factor, exemplified by the dark blue points.

\subsection{Input data preparation}\label{subsec:training}
\begin{figure}[!htbp]
\centering
\includegraphics[width=0.4\textwidth]
{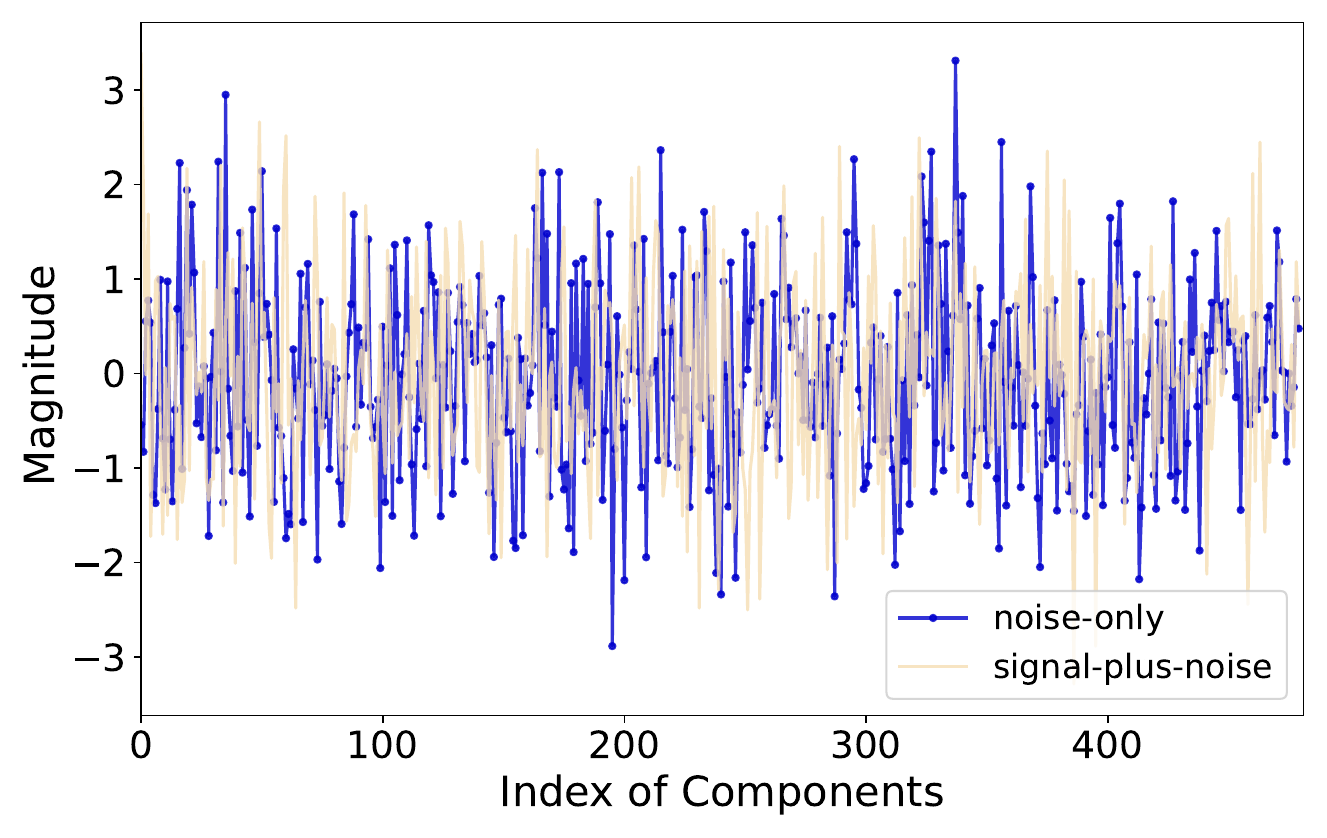}
\caption{
The projected values from two different types of input samples in the `A' channel have been normalized. 
The \ac{SNR} is set to 50. 
The associated masses of the bodies are $32.4 M_{\odot}$ and $8.62 M_{\odot}$, and they occupy a sky position marked by $\lambda = 0.17 $ Rad and $\beta = -5.17 $ Rad. Their spins are quantified as -0.65 and -0.36, while the coalescence time is set as 30020992 seconds.
A similar scenario is observed with projection values from data on the `E' channel.
}
\label{fig:eigenvalues}
\end{figure}

For both signal detection and point parameter estimation, the input data fed to the \ac{CNN} comprises amplitude projections from the \ac{IPCA}. The data from channels A and E undergo transformation using the respective \ac{IPCA} components to obtain the projection values for each sample.
The neural network performs better when the input data is normalized, shown in Figure \ref{fig:eigenvalues}. We normalized the input samples by the mean and standard deviation from the training samples.
Visually identifying a gravitational wave signal submerged in the detector noise using the naked eye presents a significant challenge.





For the signal detection task, the training data are derived from the detector channel data and consist of two distinct types: {signal-plus-noise} samples and pure noise samples. We utilize the amplitude projection values as inputs to the \ac{CNN}. Within these input samples, we assign a label of 1 if a signal is present in the original data; conversely, if the sample does not contain a signal, it receives a label of 0.

In the point parameter estimation task, the input data for the \ac{CNN} consists of amplitude projections, but exclusively from signal-plus-noise samples. These samples are obtained from simulated detector data containing signals, where the data is represented as $d(\theta) = s(\theta) + n$. To facilitate proper training, we employ normalized physical parameters as labels, as defined by the following equation:
\begin{equation}
\mathcal{{\theta}}_{\rm norm} = \frac{(\rm \mathcal{\theta}_{\rm origin} - \rm \mathcal{\theta}_{ min} ) }{(\mathcal{\theta}_{\rm max} -\mathcal{\theta}_{\rm min})}
\end{equation}
Here, $\mathcal{\theta}_{\rm origin}$ represents the original parameter value, and $\mathcal{\theta}_{\rm norm}$ serves as the normalized parameter value, which is utilized as the input label for the point estimation CNN model. 
%
During the training process for point parameter estimation, we conducted experiments with multidimensional parameter labels for the CNN. 
In comparison to using single labels for chirp mass, the utilization of multidimensional labels offers a wealth of information for training, thereby enhancing the effectiveness of \ac{CNN} training.
By experiments, we determined that 6-dimensional labels ($\mathcal{M}_{\rm norm}$, $t_{c,\rm norm}$, $\chi_{1z,\rm norm}$, $\chi_{2z,\rm norm}$, $\eta_{\rm norm}$ and $f_{0,\rm norm}$) are the most optimal choice. 

\subsection{Training}



\begin{table*}[ht]\footnotesize
\caption{\label{tab:hyperparams}The training configuration for the final experiment.
The total dataset will be divided, with 10\% allocated for validation data and the remaining 90\% used for training.}
\centering
\begin{ruledtabular}\label{process}
\begin{tabular}{llll}
        & \textbf{ setting } & \textbf{task 1 : signal detection }  &\textbf{task 2 : point estimation}  \\
\hline
\multirow{2}{*}{data samples} & &two categories:  & {signal-plus-noise samples} \\
         &        &  {signal-plus-noise} samples and noise-only samples &   \\  
        & $\rm {N_{total}}$ 
        & 12000000 
        &  12000000 
        \\
        & labels & {signals-plus-noise}: 1, {pure-noise} : 0 &  $\theta_{\rm norm}$\\
        best epoch / maximum epoch number & 
        & 63/150 
        & 149/150 
        \\
        batch size & 
        & 5000 
        & 5000 
        \\
        learning rate  &$\gamma=1 \times 10^{-5}$ & & \\
        optimizer &  Adam &   &   \\ 
        training time &   & $\sim$ 50 hours 
        &  $\sim$ 50 hours 
        \\
         \multirow{2}{*}{output model}  &  &
        detection \ac{CNN} model trained&  point estimation \ac{CNN} model  trained\\ 
        &  &   with uniform SNR [30, 50] samples &
          with the uniform SNR [30, 50] samples
\end{tabular}
\end{ruledtabular}
\end{table*}



A concise overview of  the training settings used in each task is presented in Table \ref{tab:hyperparams}, and we will delve into a more detailed explanation later.
The training strategy for both tasks is the same, therefore, let's consider the training process in task 1 (detection), as an example. The trained network is aimed at distinguishing the amplitude projections between `signal-plus-noise' samples and `noise-only' samples. It can then make judgments regarding the presence of a signal based on the projection values obtained from IPCA. 

\begin{figure}
\centering
\subfigure[]{
\begin{minipage}[b]{0.42\textwidth}
\includegraphics[width=1\textwidth]
{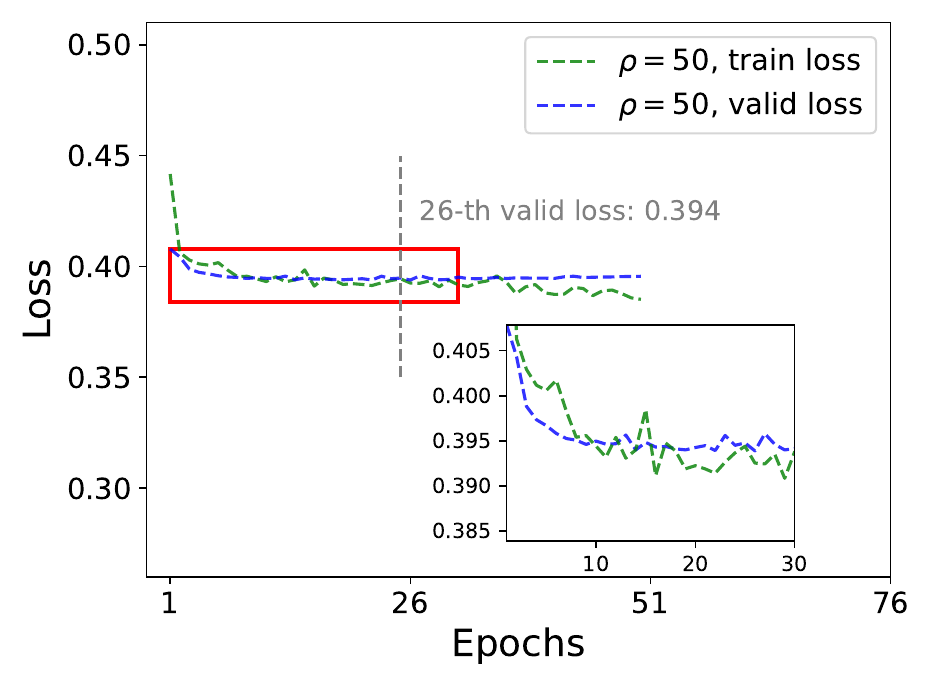} \\
\end{minipage}\label{fig:loss-bce-snr50}
}
\subfigure[]{
\begin{minipage}[b]{0.45\textwidth}
\includegraphics[width=0.9\textwidth]{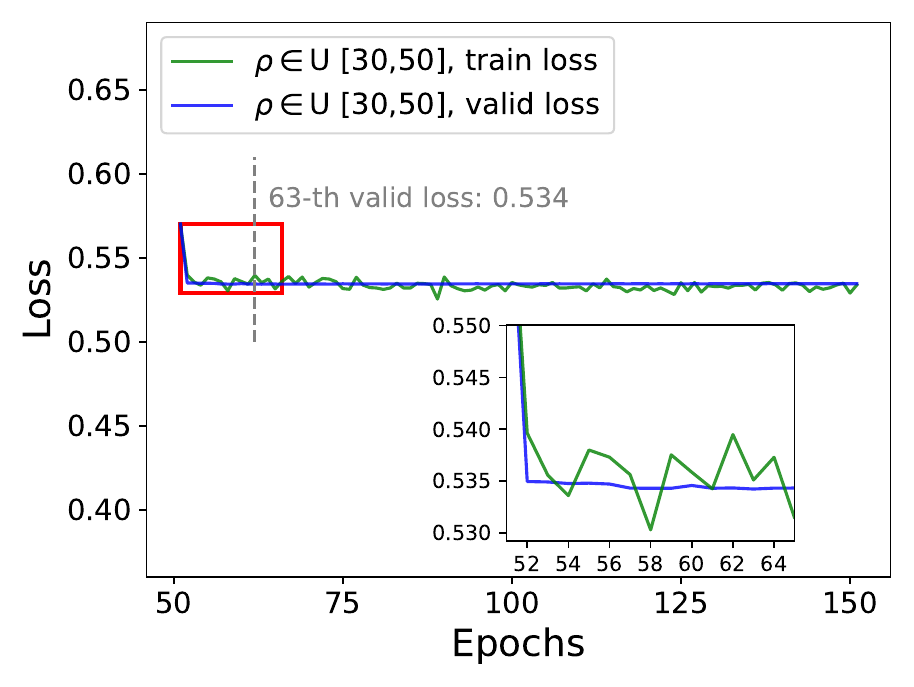} \\
\end{minipage}\label{fig:loss-bce-snr30}
}
\caption{The binary cross-entropy loss was calculated on two separate datasets for the signal detection task. The small plot insert in each subfigure showcases a magnified view of changes in the loss, marked by a red rectangle. In subfigure \ref{fig:loss-bce-snr50}, initial training was conducted over 50 epochs on a dataset with a SNR of 50, signified by dashed lines. The peak model performance, highlighted by a dotted grey vertical line, occurred at the epoch 26, serving as the pre-trained model for the succeeding dataset. In the subfigure \ref{fig:loss-bce-snr30}. The model was fine-tuned using a second data set with SNR ranging from 30 to 50. The model's optimal results surfaced at the 63rd epoch, out of 150, depicted by a dash-dotted grey vertical line.
}\label{fig:loss-bce}
\end{figure}

We start by training a \ac{CNN} model with the training datasets of the fixed \ac{SNR} of 50. 
We employed batch input with a size of 1000 to feed training samples into the network. Subsequently, we computed the binary cross-entropy loss using Equation \ref{eq:bce_loss} for each epoch.
The model's hyperparameters were then updated using the Adam optimization algorithm.  
This entire training process spans multiple epochs. 
Following each training epoch, we assess the model's performance using a validation dataset to track the training progress and mitigate overfitting issues through comparison of the validation loss with the training loss,
 such as 
 depicted in Figure \ref{fig:loss-bce}.  In response to the validation outcomes, we fine-tune the model's hyperparameters, including adjusting the learning rate and modifying the CNN architecture, to enhance its performance. Once the model has reached convergence, we decrease the fixed \ac{SNR} 50 for generating new training samples. 
 We decided to retain the working architecture for \ac{SNR} values between 30 and 50. We have chosen 30 as the lower bound for the \ac{SNR} because the network did not perform well for the lower \ac{SNR} values.

\begin{figure}
\centering
\subfigure[]{
\begin{minipage}[b]{0.42\textwidth}
\includegraphics[width=1\textwidth]
{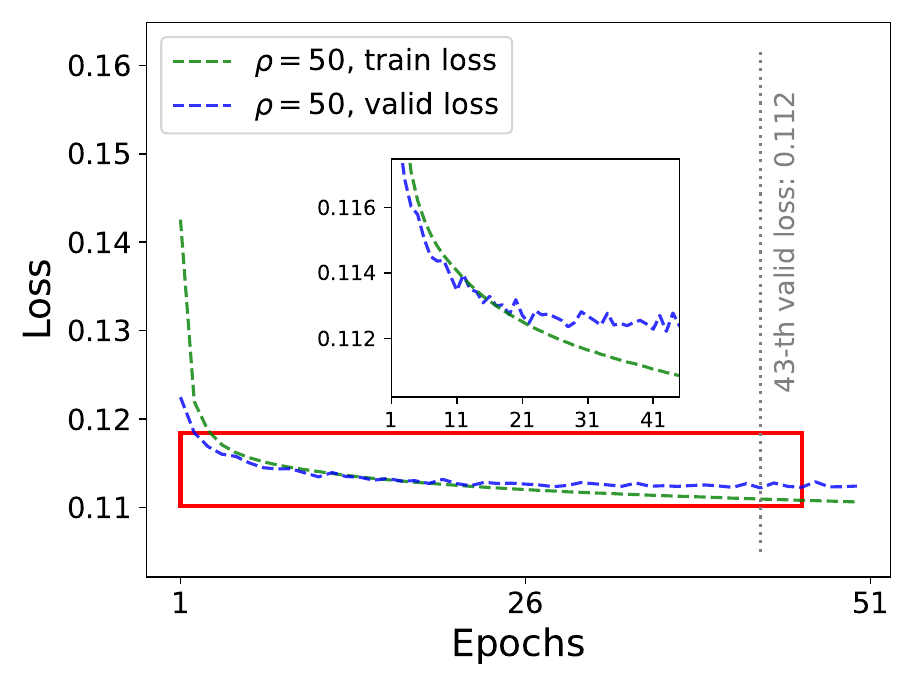} \\
\end{minipage}\label{fig:loss-mse-snr50}
}
\subfigure[]{
\begin{minipage}[b]{0.45\textwidth}
\includegraphics[width=0.9\textwidth]{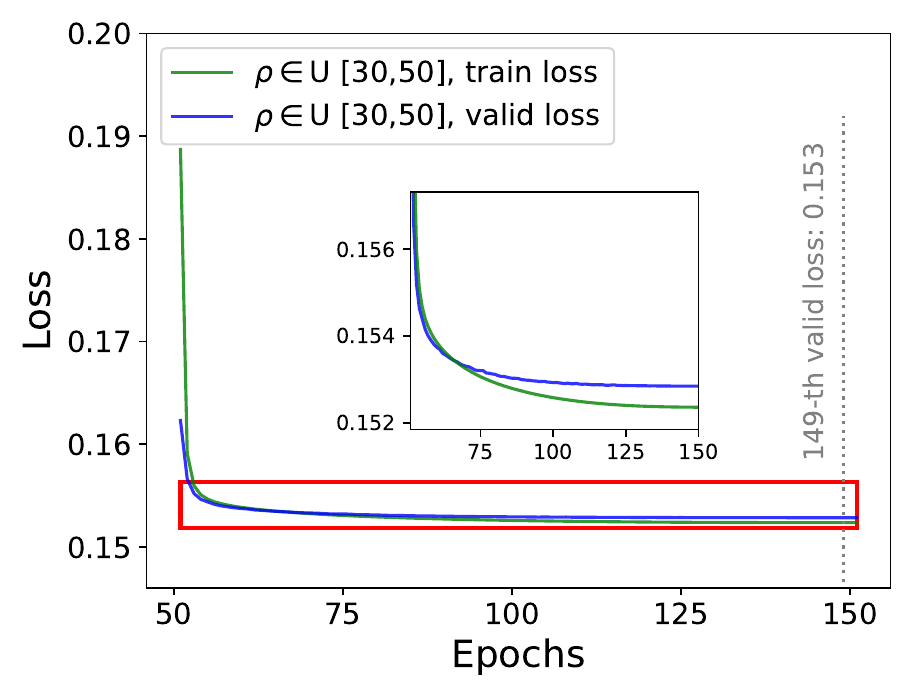} \\
\end{minipage}\label{fig:loss-mse-snr30}
}
\caption{The mean square loss was calculated on two separate datasets for the point parameter estimation task. The small plot insert in each subfigure showcases a magnified view of changes in the loss, marked by a red rectangle. In subfigure \ref{fig:loss-mse-snr50}, the first dataset had an SNR of 50, and the optimal model was achieved at the 43rd epoch, indicated by a dotted grey vertical line. This model also served as the pre-trained model for the subsequent dataset. In subfigure \ref{fig:loss-mse-snr30}, the second dataset, with an SNR varying between 30 and 50, yielded its best model at the 149th epoch out of a total of 150, as represented by a dash-dotted grey vertical line.  
}
\label{fig:loss-mse}
\end{figure}


The fine-tuned architecture of the CNN is detailed in Table \ref{tab:CNN_architecture}, consisting of a total of 6 convolution layers, 1 batch normalization  layer,  2 fully connected layers and 1 dropout layer. We configured the Adam optimizer  \cite{adam2014} with the following parameters: a learning rate of $\gamma=10^{-4}$, momentum settings of $\beta_1 = 0.9$ and $\beta_2 = 0.999$, along with a decay factor of $\epsilon = 10^{-8}$. Additional training settings are provided in Table \ref{tab:hyperparams}.
In the accompanying Figure \ref{fig:eigenvalues}, we observe that the difference in projection values between a `signal-plus-noise' sample and a `noise-only' sample is minimal, with the projection values exhibiting variations in different eigenvectors across a broad range. The introduction of a batch normalization layer after the first convolution layer plays a crucial role in data normalization, aiding the model in achieving quicker convergence.
Furthermore, we employed a learning rate reduction strategy, whereby we reduced the learning rate by a factor of 0.35 after 20 epochs in the absence of a minimum 0.005 improvement in the validation loss during that period.

In the final experiment, we utilize the top-performing \ac{CNN} with $10,000,000$ SNR 50 training samples as the pre-training model , setting an initial learning rate of 1e-4.
We then employed an additional $12,000,000$ training samples with uniformly sampled SNR ranging from 30 to 50, to continue training the model for next 100 epochs with an initial learning rate of 1e-5. This process yields a detection CNN model that converges without overfitting.
The loss curves for the uniformly sampled SNR data are shown in Figure \ref{fig:loss-bce}. The validation loss aligns closely with the training loss and keeps stable.

In task 2 (point estimation of the physical parameter ), the training strategies employed are identical to those utilized in task 1. This was decided to ensure the convolution kernels in both tasks of signal detection and point estimation would capture the same area of receptive field pertinent to \ac{sBBH} \ac{GW} signals. However, a smoother latent space was maintained in point estimation as compared to signal detection, due to the different activation functions in use. 
In addition, we only used the first dropout layer to avoid overfitting when training the CNN model with the fixed SNR 50 samples and then removed this dropout layer when training samples with an SNR range from 30 to 50. In order to train efficiently, we switched the learning rate decay method when training with the second dataset. In this dataset, we reduced it with a Cosine scheduler over 100 epochs. The initial learning rate was $1 \times 10^{-5}$ and the minimum learning rate was $1 \times 10^{-7}$.
The loss curves, displayed in Figure \ref{fig:loss-mse} , indicate that the model has been properly trained.

\section{Results}\label{sec:results}

\subsection{Detection}



We use the \ac{ROC} to describe the performance of the trained \ac{CNN} in detecting sBBH signals. When we set a detection statistical threshold \( y_{\rm {threshold}} \), if the statistic returned by the \ac{CNN} (i.e., the probability that the data when the \ac{GW} signal is present) exceeds the threshold \( y_{\rm {threshold}} \), it can be recognized as the data contains a signal; otherwise, it is  recognized as a noise sample. 
We computed both the \ac{FAP} and \ac{TAP} across various detection thresholds, constructing the \ac{ROC} curve. 
The \ac{FAP} provides a measure of the number of noise samples inaccurately identified as \ac{GW} signals. Simultaneously, the \ac{TAP}, offers a measure of the number of signal-plus-noise samples correctly identified as GW signals.

To evaluate the performance of the detection CNN model, we assemble test data with a fixed \ac{SNR}  by re-scaling the luminosity distance. Meanwhile, other source parameters remain consistent with the distribution found in the training data, as indicated in Table \ref{tab:IMRPheD-parameters-eng}. Test samples are generated at fixed \ac{SNR} of 20, 30, 40, and 50. 
We choose a reference value for false alarm probability to be 10\% because given the long duration of the signals, from several months to years, coupled with the data segment's length of three months, it's acceptable for the false alarm probability to hit one false alarm per 100 tests.
As such, every test dataset includes 2500 pure noise samples alongside an equivalent number of signal-plus-noise samples.

\begin{figure}[!ht]
\centering
\includegraphics[width=0.45\textwidth]
{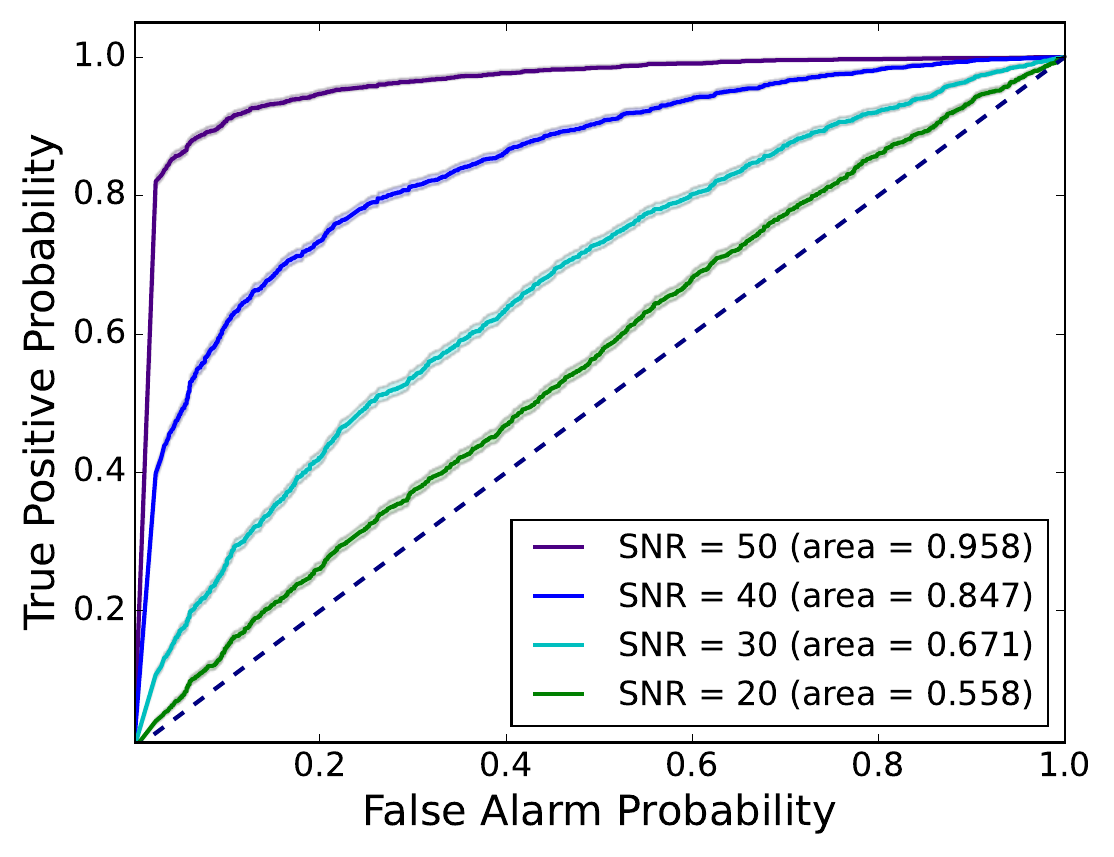}
\caption{
The ROC curves are displayed using testing data at different fixed SNR. Each curve is represented by a distinct color corresponding to a specific SNR. The \ac{AUC} quantifies the volume beneath the ROC curve, with a higher \ac{AUC} indicating a higher detection probability. Additionally, we've indicated the 1-$\sigma$ confidence intervals with shaded regions.}
\label{fig:ROC}
\end{figure}

Figure \ref{fig:ROC} showcases the detection ability of the CNN model, which has been trained with \ac{SNR} uniformly distributed within the 30 to 50 range, to differentiate signals based on their varying \ac{SNR}. It is observed that an increase in the \ac{SNR}  corresponds to an augmentation in the \ac{TAP}, with the specifics delineated in Table \ref{tab:TAP}.  The final trained detection CNN model can effectively detect \ac{sBBH} sources with an SNR of 50 or higher, achieving an area under the curve (AUC)  of 0.958.  Additionally, it  demonstrates that sources with fixed \acp{SNR} of 40, 30, and 20 achieved corresponding \acp{AUC} values of 0.847, 0.671, and 0.558, respectively.


\begin{table}\label{tab:TAP}
\caption{When \ac{FAP} is set to 10\% , the corresponding \ac{TAP} for each distinct \ac{SNR} testing data set are individually displayed refs to the Figure. \ref{fig:ROC} }
\begin{tabular}{ccccc}
\hline
\hline
\multirow{2}{*}{FAP = 10\%} 
 & \multicolumn{4}{c}{SNR}  \\
\cline{2-5}
 & 50 & 40 & 30 & 20  \\
\hline
 TAP (\%)&91.0 &61.7 & 27.3 &14.9 \\
\hline
\hline
\end{tabular}
\end{table}





\subsection{Mass estimation}

\begin{figure}[!ht]
\centering
\includegraphics[width=0.5\textwidth]%
{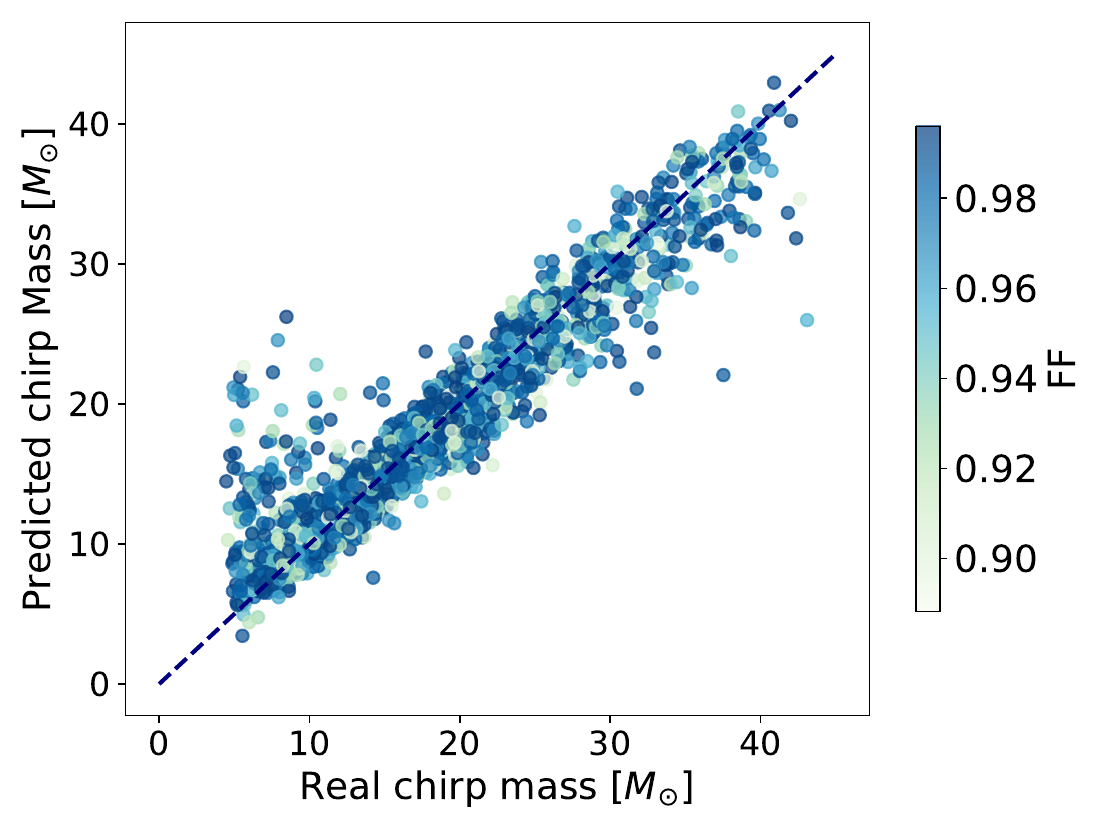}
\caption{
The scatter plot illustrates the estimation of chirp mass of the fixed SNR 50 test signals compared to the actual values. The horizontal axis represents the actual chirp mass of the test samples, while the vertical axis displays the predictions obtained from the point estimation \ac{CNN} model trained with uniformly sampled SNR samples.
The error is calculated by subtracting the predicted chirp mass from the actual chirp mass: $\rm error = \mathcal{M}_{\rm actual} - \mathcal{M}_{\rm pred}$.
}
\label{fig:prediction}
\end{figure}

In order to estimate the value of the chirp mass by the point parameter estimation \ac{CNN} model trained with a uniformly sampled SNR ranging from 30 to 50, we took 2000 test samples with a fixed \ac{SNR} of 50 which had the same parameter distribution as the training samples. As shown in Figure \ref{fig:prediction}, most of the chirp mass predictions are tightly distributed along the diagonal, indicating that there is consistency between the predicted values and actual values of the source's chirp mass.
In this test dataset, 90\% of the samples have an absolute error of 2.49 $M_{\odot}$ and a relative error of 0.13. 
This means that our model can predict the chirp mass of SNR 50 signals in TianQin runs.
Leaving aside those sources with larger errors, the significant variance observed in most sources yielding accurate predictions can largely be ascribed to added detector noise. 

We also studied the characteristics of signals with larger prediction errors for chirp mass (absolute error exceeding 10 solar masses). We found that some signals have a smaller fitting factor, indicating that more information was lost after the \ac{IPCA} projection. 
Signals characterized by a larger symmetric mass ratio 
will exhibit a reduced number of data points in frequency domain within the same duration. 
This particular property may introduce challenges for \acp{CNN} as they strive to effectively learn these signal features and make accurate predictions of the chirp mass. 
From the current results, the point estimation \ac{CNN} can still quickly provide chirp mass predictions for most \ac{sBBH} signals of SNR 50 with a good accuracy.

\section{Conclusion and discussion}\label{sec:conclusion}

In this study, we effectively compressed the data through low rank matrix approximation, and extracted the main features of the frequency signal amplitude, thereby building the \ac{IPCA} model. By projecting the simulated detector data (including signal plus noise samples and pure noise samples) through \ac{IPCA}, we classify the compressed data into two categories by a trained \ac{CNN} and show the sensitivity via the ROC curve. Simultaneously, in terms of the point estimation of parameters, we utilized the \ac{CNN} model to estimate chirp mass, marginalising over all other parameters, for the detector data containing sBBH signals with a SNR of 50, 
and estimated its measurement error (90\% confidence interval) to be 2.49 solar masses. 

Regarding \ac{sBBH} search performance, we have also recognized the potential for further enhancing our search model's sensitivity. 
Given the challenges of analyzing  prolonged  \ac{GW} signals  and the information loss incurred during the \ac{IPCA} data compression process, the priority lies in improving the compression algorithm and feature extraction rather than further enhancing the existing neural network architecture. To be specific, the current \ac{IPCA} model demonstrates certain limitations in reconstructing complex signals, there remains a need to investigate the influence of wave source parameter distribution and sampling methods on \ac{IPCA} performance.  Therefore, in future research, we plan to explore feature extraction methods aimed at minimizing information loss while maintaining high compression efficiency. These may include using feedforward neural networks, as proposed in \cite{2023arXiv230716668G}.

Additionally, some might suggest utilizing a classic architecture of CNN.
We attempted to use the $\textit{resnet-18}$ model (with approximately $10^7$ hyperparameters), but it exhibited susceptibility to overfitting and required a significantly larger number of training samples and training time. Specifically, our $\textit{resnet-18}$ experiment demanded around $10^7$ training samples and approximately 300 hours for 100 training epochs, which was more than three times the computational time of our current model. 

Looking at the analysis of \ac{sBBH} search results using the TianQin detector, according to existing theoretical analyses \cite{Moore:2019pke}, confirming the presence of signals in data from a single space-borne detector is a formidable challenge. 
The challenge arises primarily from the immense number of templates required for a coherent search, exceeding $10^{31}$ \cite{Moore:2019pke}, which makes confirming the presence of signals within data from a single space-borne detector particularly difficult.
Despite the absence of phase information in our study and the information loss during \ac{IPCA}, the neural networks can still effectively detect some \ac{sBBH} signals at a 10\% false alarm probability. If a \ac{sBBH} candidate is proximate to our observatory, our detection \ac{CNN} model is highly likely to identify it with relatively little computational burden.
For future \ac{sBBH} signal detection, we consider the joint observation by multiple detectors, such as TianQin and LISA, with the hope of lowering the \ac{SNR} threshold, providing richer information for the neural network, and effectively detecting \ac{sBBH} signals with lower \ac{SNR}. Additionally, we may explore the use of more complex neural network architectures to enhance sensitivity.

In terms of point estimation of chirp mass, the current \ac{CNN} results demonstrate the ability of neural network algorithms to  capture the characteristics of \ac{sBBH} signals, although achieving accurate prediction for all physical parameters has proved challenging in our experiments. This is largely due to our reliance on amplitude in the frequency domain. One possible improvement could involve expanding the information input into the neural network, including data and label representations. Furthermore, optimizing the neural network architecture used, such as considering the application of autoencoders, and normalizing flow, can contribute to more precise parameter estimation.

From a broad perspective, the sensitivity of our pipeline is more efficient than the excess power method, which only uses amplitude. This is because we incorporate additional signal information during compression. The extracted bases encapsulate the correlation of data points in signals - if the detector records data in the presence of a signal, this correlation becomes evident during compression. This compression technique could also enhance our understanding of the latent space for \ac{sBBH} signals or it could be employed to assess the reduced bias across the entire template bank, as demonstrated in the paper \cite{Field:2011mf}. Moreover, this strategy of combining data compression and search presents a promising new approach for the detection of long-lived sBBH signals. Intending to refine this method further in the future, our search for early inspiral phases of sBBH will directly contribute to a better understanding of the formation channels of sBBH.

\begin{acknowledgments}

X. Z. would like to express profound gratitude toward En-Kun Li, Xiangyu Lyu, Han Wang, Shuai Liu, and Jian-dong
Zhang for insightful input at the initial stage of this project.
Special thanks are also extended to Naren Nagarajan at the
University of Glasgow for training discussions. 
Y. H. is supported by the Natural Science Foundation of China
(Grants No. 12173104, No. 12261131504), the National
Key Research and Development Program of China
(No. 2023YFC2206700, No. 2020YFC2201400), and
Guangdong Major Project of Basic and Applied Basic
Research (Grant No. 2019B030302001). X. Z. is supported
by CSC and the Royal Society -  IEC\textbackslash NSFC \textbackslash 211371.
\end{acknowledgments}

\appendix

\section{IPCA}

\begin{figure}[!ht]
\centering
\includegraphics[width=0.45\textwidth]{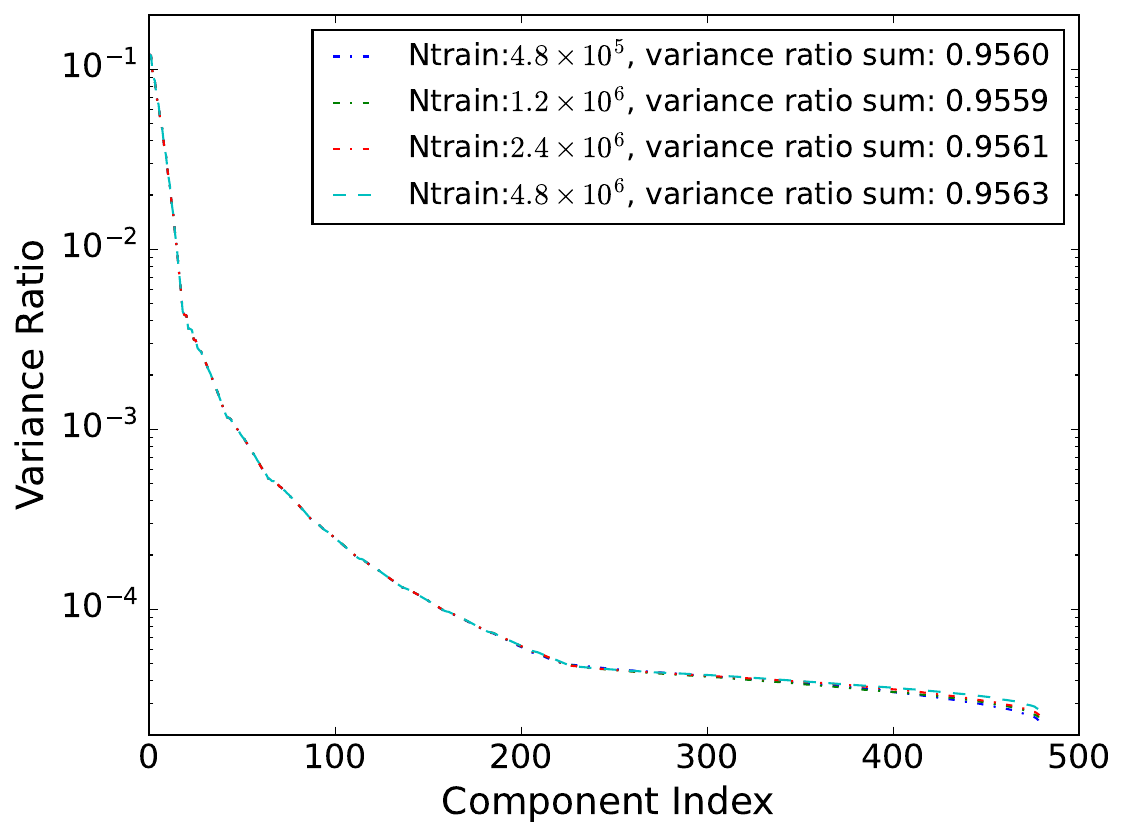}
\caption{Variance ratio plotted against the number of training signals for the IPCA$^A$ model. }
\label{fig:ipca-Ntrain}
\end{figure}

We extracted 480 basis for the IPCA models, constrained by the processing capacity of our computational cluster, notably the GPU memory. This appendix was done to demonstrate the influence of the number of training signals when constructing a compression model using IPCA.  In the figure \ref{fig:ipca-Ntrain}, the variance ratio for different IPCA models within the A-channel signal exhibits minimal change. Concurrently, the basis also registered minor adjustments.

\bibliography{apssamp}

\end{document}